\documentclass[]{aa}
\usepackage{hyperref}
\usepackage[capitalise]{cleveref} 
\usepackage{morefloats}
\usepackage{amsmath}
\usepackage{color}
\usepackage{textcomp}
\usepackage{commath}

\bibpunct{(}{)}{;}{a}{}{,} 

\begin{document}

\title{The Effects of Consistent Chemical Kinetics Calculations on the Pressure--Temperature Profiles and Emission Spectra of Hot Jupiters}

\titlerunning{The Effects of Consistent Chemical Kinetics Calculations in Hot Jupiters}
\authorrunning{Drummond et al.}

\author{B. Drummond\inst{\ref{inst1}}
\and P. Tremblin\inst{\ref{inst1},\ref{inst2}}
\and I. Baraffe\inst{\ref{inst1},\ref{inst5}}
\and D.S. Amundsen\inst{\ref{inst1},\ref{inst3},\ref{inst4}}
\and N. J. Mayne\inst{\ref{inst1}}
\and O. Venot\inst{\ref{inst6}}
\and J. Goyal\inst{\ref{inst1}}}

\institute{Astrophysics Group, University of Exeter, EX4 4QL, Exeter, UK\label{inst1}  \\ 
\email{bdrummond@astro.ex.ac.uk}
\and Maison de la Simulation, CEA-CNRS-INRIA-UPS-UVSQ, USR 3441 Centre d'\'{e}tude de Saclay, 91191 Gif-Sur-Yvette, France\label{inst2}
\and Department of Applied Physics and Applied Mathematics, Columbia University, New York, NY 10025, USA\label{inst3}
\and NASA Goddard Institute for Space Studies, New York, NY 10025, USA\label{inst4}
\and Univ Lyon, Ens de Lyon, Univ Lyon1, CNRS, CRAL, UMR5574, F-69007, Lyon, France\label{inst5}
\and Intituut voor Sterrenkunde, Katholieke Universiteit Leuven, Celestijnenlaan 200D, 3001 Leuven, Belgium\label{inst6}}

\date{Received /
	Accepted }


\abstract {In this work we investigate the impact of calculating non-equilibrium chemical abundances consistently with the temperature structure for the atmospheres of highly--irradiated, close--in gas giant exoplanets. Chemical kinetics models have been widely used in the literature to investigate the chemical compositions of hot Jupiter atmospheres which are expected to be driven away from chemical equilibrium via processes such as vertical mixing and photochemistry. All of these models have so far used pressure--temperature ($P$--$T$) profiles as fixed model input. This results in a decoupling of the chemistry from the radiative and thermal properties of the atmosphere, despite the fact that in nature they are intricately linked. We use a one-dimensional radiative-convective equilibrium model, \texttt{ATMO}, which includes a sophisticated chemistry scheme to calculate $P$--$T$ profiles which are fully consistent with non-equilibrium chemical abundances, including vertical mixing and photochemistry. Our primary conclusion is that, in cases of strong chemical disequilibrium, consistent calculations can lead to differences in the $P$--$T$ profile of up to 100 K compared to the $P$--$T$ profile derived assuming chemical equilibrium. This temperature change can, in turn, have important consequences for the chemical abundances themselves as well as for the simulated emission spectra. In particular, we find that performing the chemical kinetics calculation consistently can reduce the overall impact of non-equilibrium chemistry on the observable emission spectrum of hot Jupiters. Simulated observations derived from non-consistent models could thus yield the wrong interpretation. We show that this behaviour is due to the non-consistent models violating the energy budget balance of the atmosphere.}

\maketitle

\section{Introduction}

Despite the discovery of ever smaller and more Earth-like rocky exoplanets \citep[e.g.][]{Berta-Thompson2015}, hot Jupiters remain one of the most important classes of exoplanet due to the availability of follow-up characterisation observations. This allows for important comparisons between the various observables (transmission spectra \citep[e.g.][]{Sing2016}, emission spectra \citep[e.g.][]{Knutson2008,Beaulieu2010,Diamond-Lowe2014,Evans2015} and phase curves \citep[e.g.][]{Zellem2014}; see \citet{HengShowman2015} for a review) with the many atmosphere models in use throughout the community to understand the physical and chemical processes occuring in these atmospheres.

There exists a hierarchy of atmosphere models which have so far been applied to the study of exoplanets with each class of model having its own practical use. The three--dimensional (3D) general circulation models (GCMs) are required to gain insights into the atmospheric dynamics \citep[e.g.][]{Showman2009,Burrows2010,Heng2011,RauscherMenou2012,Dobbs-Dixon2013,Mayne2014,Polichtchouk2014} whilst the one--dimensional (1D) radiative--convective models solve for single column pressure--temperature ($P$--$T$) profiles \citep[e.g.][]{Iro2005,Barman2005,Fortney2010,SpiegelBurrows2010}. A further class of atmosphere model, the chemical kinetics models, determine precise chemical compositions \citep[e.g.][]{Moses2011,Venot2012,Zahnle2014}. 

Early work using these chemical kinetics models focussed on the photochemical production and destruction processes of hydrocarbon hazes \citep{Liang2004,Zahnle2009a} and the production of atomic hydrogen through water photolysis \citep{Liang2003}. Later modelling efforts provided detailed chemical compositions of these atmospheres and considered the impact on the observable properties \citep{Moses2011,Venot2012,Madhusudhan2011}.

The non-equilibrium processes of vertical mixing and photochemistry have been shown to have notable consequences on the simulated spectra of some hot exoplanet atmospheres \citep[see][for a review]{Moses2014}. Vertical mixing can result in transport-induced quenching which increases or decreases the abundances of chemical species compared with their chemical equilibrium profiles. In addition, at low pressures high-energy photons can dissociate molecules into highly reactive daughter products.

These non-equilibrium processes have the potential to influence the simulated spectrum of the atmosphere by changing the amount of absorption or emission from these molecules. For example, \citet{Moses2011} found that increases in the abundances of methane and ammonia due to transport-induced quenching have important consequences for the emission spectrum of HD 189733b, decreasing the calculated eclipse depth over several wavelength regions.

One common feature amongst all of the chemical kinetics studies published so far is that none include a consistent approach to the model atmosphere. The background thermal structure of the atmosphere is treated as a fixed model input. Usually the supplied $P$--$T$ profiles have been derived from 3D GCM results \citep[e.g.][]{Moses2011,Venot2012} or 1D radiative-convective models \citep[e.g.][]{Moses2013,Agundez2014b}. The equilibrium and non-equilibrium chemical abundances are then calculated using this fixed $P$--$T$ profile leading to an inconsistency between the chemical abundances and the thermal structure. As the chemistry is driven away from chemical equilibrium, the opacity in the model atmosphere changes which should impact on the $P$--$T$ profile.

\citet{Agundez2014b} have attempted to take into account this effect by including one additional iteration in their radiative-convective model using the initial non-equilibrium chemical abundances they calculate for the hot Neptune GJ 436b. They found corrections to the temperature structure of $<$ 100 K. In addition, \citet{Hubeny2007} performed consistent non-equilibrium chemistry models of brown dwarf atmospheres using a timescale argument to quench the abundances of N$_2$/NH$_3$ and CO/CH$_4$, not using chemical kinetics, and found corrections to the $P$--$T$ profile of 50--100 K. 

In this study we use a 1D atmosphere code which solves for both hydrostatic and energy balance and includes a sophisticated chemistry scheme to solve for the non-equilibrium chemical abundances, including vertical mixing and photochemistry, consistently with the $P$--$T$ profile. In \cref{section:model_description} we describe our model setup. In \cref{section:results} we present our results and, finally, in \cref{section:conclusion} we summarise and conclude.

\section{Method and Model Description}
\label{section:model_description}

We use the 1D atmosphere code \texttt{ATMO} \citep{Amundsen2014,Tremblin2015,Tremblin2016} to generate models of two highly irradiated exoplanet atmospheres. The model has already been applied to the atmospheres of brown dwarfs \citep{Tremblin2015,Tremblin2016} and we use essentially the same model setup here except that external irradiation is now included. In this section we present details of the model and our method for calculating non-equilibrium chemical abundances consistently with the $P$--$T$ profile.

\subsection{Radiative-Convective Equilibrium}
\texttt{ATMO} solves for the pressure-temperature structure of an atmosphere by finding energy flux balance in each model level, i.e.,
\begin{equation}
\label{equation:energy}
  \int_0^\infty\left(F_\mathrm{rad}(\nu) +F^0_\mathrm{star}(\nu)e^{\tau_\nu/\mu_\mathrm{star}}\right)d\nu + F_\mathrm{conv} = \sigma T_\mathrm{int}^4
  \end{equation}
where $F_\mathrm{rad}(\nu)$ and
$F_\mathrm{conv}$ are the spectral radiative flux and the convective flux,
respectively, and $\tau_\nu$ is the vertical monochromatic optical depth. $T_\mathrm{int}$ is the internal temperature of the object
corresponding to the surface flux at which the object cools in the
absence of irradiation; $T_\mathrm{int}$ is equivalant to the effective temperature $T_\mathrm{eff}$ in the absence of irradiation. $F^0_\mathrm{star}(\nu)$ is the incoming
irradiation flux from the star at the top of the atmosphere and
$\mu_\mathrm{star}$ = $\cos\theta$ where $\theta$ is the angle of incoming radiation off the vertical; $\mu_\mathrm{star}$ is negative. In addition, the
 pressure as a function of altitude is calculated using the equation of hydrostatic balance
\begin{equation}\label{eq:hydrostat}
 \frac{\partial P}{\partial z}  = -\rho g
  \end{equation}
where $P$ is pressure, $z$ is altitude, $\rho$ is density
and $g$ is gravity. An object's atmosphere is therefore completely determined
by its internal temperature and gravity, often specified as
$(T_\mathrm{int}, \log(g))$, in the absence of irradiation. The
convective flux is computed using the mixing length theory as
presented by \citet{Henyey1965}
\begin{equation}
F_\mathrm{conv} = \frac{1}{2} \rho C_p T v_\mathrm{conv}
\frac{l}{H_p}\frac{\Gamma}{\Gamma+1}\left(\nabla_T - \nabla_\mathrm{ad}\right)
\end{equation}
with $v_\mathrm{conv}$ the convective velocity, $\Gamma$ the efficiency parameter, $\nabla_T=\partial \log(T)/\partial \log(P)$,
$\nabla_\mathrm{ad}$ the adiabatic gradient, $C_p$ the specific heat
at constant pressure, $H_p$ the pressure scale height, and $l = \alpha H_p$ the
mixing length. The details of the 
computation of $v_\mathrm{conv}$ and $\Gamma$ can be found in
\citet{Gustafsson2008} and we used the same standard parameters
(temperature distribution in a convective element $y=0.076$, energy
dissipation by turbulent viscosity $\nu = 8$, mixing length\footnote{The choice of $\alpha$ is based on typical values, namely $\sim$ 1.5 - 2,  used in cool object atmosphere models (low mass stars and brown dwarfs) and based on calibration of 1D models with multi-D radiative hydrodynamic 
simulations \citep[e.g.][]{Baraffe2015}. Such calibration has never been performed for hot Jupiter-like planets 
but we note that present hot Jupiter models are rather insensitive to this parameter for any value $\ge$ 1.5.} $\alpha =
1.5$). The non-linear problem given by Eq.~\ref{equation:energy} and
Eq.~\ref{eq:hydrostat} is solved by a Newton-Raphson algorithm to
obtain the converged $P$--$T$ profile of the atmosphere. 

\subsection{Radiative Transfer}
\label{section:radiative_transfer}

The radiative transfer equation is solved in 1D plane-parallel
geometry including isotropic scattering.  
We include CH$_4$, H$_2$O, CO, CO$_2$, NH$_3$,
TiO, VO, Na, K, Li, Cs and Rb and collision induced absorption due to H$_2$-H$_2$ and H$_2$-He as opacity sources in the atmosphere
\citep[see][for details]{Tremblin2015,Amundsen2014}. \texttt{ATMO} can be used as a full line-by-line code at high
spectral resolution (0.001 cm$^{-1}$, evenly spaced in wavenumber), e.g. to post process high-resolution emission and
transmission spectra, though this is not used in this study. 

Presently \texttt{ATMO} is used with the correlated-$k$
approximation using the random
overlap method to compute the total mixture opacity (see \citet{Lacis1991} and Amundsen et al. in prep) for moderate
resolution emission and transmission spectra with 500 or 5000 bands
and, for a rapid computation of the radiative flux, with 32 bands in
the Newton-Raphson iterations. The number of $k$-coefficients per band is not fixed but is instead computed based on a specified precision; the number of $k$-coefficients therefore varies depending on the band and the gas \citep[this is explained in detail in][]{Amundsen2014}. The $k$-coefficients were derived for opacity tables as in \citet{Amundsen2014}. The results obtained using the correlated-$k$ approximation have been compared with the full line-by-line result, which agree very well.

The
radiative transfer equation for the irradiation and thermal emission
of the planet is split into a directly-irradiated part
$I_\mathrm{star}(\nu)$ and a thermal
plus diffusely-irradiated part $I_\mathrm{rad}(\nu) 
I_\mathrm{therm}(\nu) + I_\mathrm{diff}(\nu)$ \citep[see][chap. 6.2 for details]{Thomas1999}. The direct part can be
integrated straightforwardly to give
\begin{equation}
  F_\mathrm{star}(\nu) =2\pi \int^1_{-1} I_\mathrm{star}(\nu)\mu d\mu = F^0_\mathrm{star}(\nu)e^{\tau_\nu/\mu_\mathrm{star}}
\end{equation}
where $F^0_\mathrm{star}(\nu)$ is the top incoming flux defined as
$F^0_\mathrm{star}(\nu) = 4\pi H^0_\mathrm{star}(\nu) \mu_\mathrm{star}
R_\mathrm{star}^2/R_\mathrm{orbit}^2$; here $H^0_\mathrm{star}(\nu)$ is the flux at the surface of the star, and $R_\mathrm{star}$ and $R_\mathrm{orbit}$ are the stellar and orbital radius respectively.  The diffuse and thermal part are computed using the radiative transfer
equation with isotropic scattering \citep{Thomas1999,HubenyMihalas2014}
\begin{eqnarray}
  \mu \frac{dI_\mathrm{rad}(\nu)}{d\tau_\nu} =
  I_\mathrm{rad}(\nu)-(1-\epsilon_\nu)J_\mathrm{rad}(\nu)- \cr
 (1-\epsilon_\nu)F_\mathrm{star}(\nu)/4\pi - \epsilon_\nu B(\nu)
\end{eqnarray}
where $\epsilon_\nu$ is the photon destruction probability given as a
function of absorption and scattering opacity
$\kappa_\mathrm{abs}/(\kappa_\mathrm{scat}+\kappa_\mathrm{abs})$,
$B(\nu)$ is the Planck function, and 
\begin{eqnarray}
  J_\mathrm{rad}(\nu) &=& \frac{1}{2}\int^1_{-1}I_\mathrm{rad}(\nu)d\mu\cr
  F_\mathrm{rad}(\nu) &=&
  2\pi\int^1_{-1}I_\mathrm{rad}(\nu)\mu d\mu
\end{eqnarray}
The ray directions specified by $\mu$ are sampled with a discrete-ordinate method
using Gauss-Legendre quadrature (we usually use 16 rays in total). We solve the radiative transfer equation iteratively in its integral form
following \citet{Bendicho1995} and using a Gauss-Seidel-type
$\Lambda$-acceleration scheme for the scattering
\begin{eqnarray}
 I_\mathrm{rad}(\nu) &=& I^0_\mathrm{rad}(\nu)e^{-\abs{({\tau_\nu/\mu})}} +
 \int_0^{\tau_\nu/\mu} S_\nu(t) e^{\tau'-\abs{(\tau_\nu/\mu)}} d\tau' \cr
 S_\nu&=& (1-\epsilon_\nu)J_\mathrm{rad}(\nu)+(1-\epsilon_\nu)F_\mathrm{star}(\nu)/4\pi +\cr
 \epsilon_\nu B(\nu)
\end{eqnarray}

The radiative-transfer scheme has been benchmarked against the Met
Office SOCRATES code \citep{Amundsen2014}. All results presented in this study include isotropic scattering.

\subsection{Chemistry: Gibbs Energy Minimisation}
\label{section:gibbs_minimisation}

The Gibbs energy minimisation scheme is one of two chemistry schemes included in the model. Minimisation of the Gibbs energy is a popular method for obtaining chemical equilibrium abundances in atmosphere models \citep{Burrows1999,Blecic2015}. This technique does not rely on complicated chemical networks and is therefore much simpler to implement and compute than a chemical kinetics method.

We follow the method of \citet{Gordon1994}, using the same thermochemical data as \citet{Venot2012} in the form of NASA polynomial coefficients \citep{McBride1993}. The Gibbs mimimisation method allows for depletion of gas phase species due to condensation, however we do not consider condensation here and this is left for a future study.

At constant temperature and pressure, the Gibbs energy of the system, $G$, which is given by
\begin{equation}
	G = \sum_{i=1}^I \mu_i n_i,
\end{equation}
where $\mu_i$ and $n_i$ are the chemical potential and number of moles of chemical species $i$, must be minimised subject to the constraint of elemental conservation,
\begin{equation}
	\sum_{i=1}^Ia_{ij}n_i - b_j^0 = 0,
\end{equation}
where $a_{ij}$ is the number of atoms of element $j$ per molecule $i$ and $b_j^0$ is the total number of moles of element $j$ in the mixture. This minimisation is performed using the method of Lagrange multipliers.

We include $\sim$140 chemical species in total, with 105 of those species being the ones contained in the \citet{Venot2012} chemical network. In addition to those, we include Na, K, Li, Cs and Rb in monatomic form and their important molecular gas-phase species (e.g. NaCl, NaOH,...); which can deplete the abundance of the monatomic forms. Local chemical equilibrium abundances are acheived by minimising the Gibbs energy independently in each model level.

We benchmark the Gibbs minimisation scheme against the chemical equilibrium analytical formulae of \citet{Burrows1999} and \citet{Heng2016} in \cref{section:chemical_gibbs_test}.

\subsection{Chemistry: Chemical Kinetics}
\label{section:chemical_kinetics}

We also include a chemical kinetics scheme which deals directly with chemical reactions. We solve for the chemical steady-state by integrating the continuity equation,
\begin{equation}
\label{equation:continuity}
\frac{\partial n_i}{\partial t} = P_i - n_iL_i - \frac{\partial \Phi_i}{\partial z},
\end{equation}
so that $\frac{\delta n_i}{\delta t}$ $\sim$ 0. Here, $P_i$ and $L_i$ are the chemical production and chemical loss terms, derived from the system of chemical reactions in the chemical network, and the $\frac{\partial \Phi_i}{\partial z}$ term describes vertical mixing. 

The vertical transport flux $\Phi_i$ is split into two components due to molecular diffusion and eddy diffusion and is described by \citep[e.g.][]{Gladstone1996},
\begin{equation} 
\label{equation:flux_n}
\begin{split}
	\Phi_i = -D_i\left(\frac{\partial n_i}{\partial z} + \frac{n_i}{H_i} + \frac{n_i(1 + \alpha)}{T}\frac{d T}{d z}\right) \\
-K_{zz}\left(\frac{\partial n_i}{\partial z} + \frac{n_i}{H_a} + \frac{n_i}{T}\frac{d T}{d z}\right).
\end{split}
\end{equation}
where $D_i$ and $K_{zz}$ are the molecular and eddy diffusion coefficients, respectively, and $H_i$ and $H_a$ are the scale heights of each individual species and of the bulk atmosphere, respectively, and $\alpha$ is the thermal diffusion parameter. \cref{equation:flux_n} can be written more simply in terms of the mole fraction $f_i$ = $n_i$/$n$;
\begin{equation}
\begin{split}
	\Phi_i = -nD_i\left(\frac{\partial f_i}{\partial z} - \frac{f_i}{H_a} + \frac{f_i}{H_i} + \frac{f_i\alpha}{T}\frac{d T}{d z}\right) \\
-nK_{zz}\left(\frac{\partial f_i}{\partial z}\right).
\end{split}
\end{equation}

The molecular diffusion coefficient $D_i$ can be determined using the kinetic theory of gases \citep[e.g.][]{Wayne1991} and is inversely proportional to the number density. Therefore, molecular diffusion becomes important at low pressures ($P$ $<$ 10$^{-5}$ bar) in the thermospheric regions of the atmosphere. At higher pressures, vertical transport is dominated by the eddy diffusion term for which the controlling coefficient $K_{zz}$ is far less well constrained, both observationally and theoretically. In previous studies, this term has been estimated using wind velocity fields derived from 3D general circulation models \citep[e.g.][]{Moses2011} or parameterised using the advection of passive tracers \citep{Parmentier2013}; however if the turbulent diffusion is caused by small scale processes linked to convection, overshooting of the convective region, gravity waves or fingering convection \citep{Tremblin2016}, then care must be taken when deriving this parameter from 3D simulations in which the distinction should be made between the large scale vertical advection linked to the jet circulation and the small scale turbulent diffusion processes that are likely to be unresolved. Other studies treat the $K_{zz}$ parameter as a free model parameter and test a range of plausible values \citep[e.g.][]{Miguel2014}.

In this work, we choose two plausible values for the $K_{zz}$ parameter and show models for both cases. Values of the $K_{zz}$ parameter for hot Jupiter atmospheres used in other works vary between $\sim$10$^{7}$--10$^{12}$ cm$^{2}$s$^{-1}$ \citep[e.g.][]{Moses2011,Miguel2014}. We take one case roughly in the middle of this range ($K_{zz}$ = 10$^{9}$ cm$^{2}$s$^{-1}$) and one towards the upper limit of this range ($K_{zz}$ = 10$^{11}$ cm$^{2}$s$^{-1}$). We impose zero-flux limits on both the upper- and lower-boundaries, assuming that no mass enters the atmosphere at the bottom or escapes the atmosphere at the top. In addition, we chose the value of $K_{zz}$ and maximum pressure level of our model such that the lower boundary will always remain in chemical equilibrium, meaning that our assumption of zero-flux at the lower boundary is valid \citep{Moses2011}.

The chemical production and loss terms, $P_i$ and $L_i$, are calculated from the system of chemical reactions contained within the chemical network employed. In this study, we adopt the \citet{Venot2012} chemical network previously applied in several studies of hot, hydrogen-dominated exoplanet atmospheres \citep{Agundez2012,Agundez2014,Agundez2014b,Venot2014} and for brown dwarf atmospheres \citep{Tremblin2015,Tremblin2016}. 

We choose to use the original C$_0$-C$_2$ network of \citet{Venot2012} with hydrocarbon species of up to two carbon atoms, rather than the more recent update which includes higher-order hydrocarbons C$_0$-C$_6$ \citep{Venot2015}. The overall aim of this work is to study the effect of non-equilibrium chemistry on the background atmosphere and it is therefore most important to obtain accurate abundances for the species contributing to the opacity. The inclusion of higher-order hydrocarbons was found not to affect the abundances of the main species (e.g. CH$_4$, H$_2$O) \citep{Venot2015}. Therefore we choose to use the smaller and less computationally expensive C$_0$--C$_2$ network.

We use the LSODE solver\footnote{https://computation.llnl.gov/casc/odepack/} \citep{Hindmarsh1983} using the DLSODES option for stiff system of ordinary differential equations to solve the system of continuity equations simultaneously (\cref{equation:continuity}) for each chemical species using the Backwards Differentiation Formulae method, until at least 10$^{12}$ s when the chemistry has reached a steady-state. There are two options to choose the timestep of the chemistry iterations: a customised version which starts with an initial timestep of $dt = 10^{-10}$ s which is progressively increased, by 10\%, if the maximum relative variations of the molecular abundances do not exceed 10\%, or we employ the timestep which is internally generated within the DLSODES routine. We have tested both methods, which give the same results for a given time. Since the timestep generated by DLSODES is typically larger than that using our custom method, we generally use the second option.

As well as vertical mixing, dissociations of molecules by energetic X--ray and UV (XUV) photons can drive the chemistry away from local chemical equilibrium at low pressures where the flux of these photons is high. The XUV photon flux is calculated as a function of pressure, using the same radiative transfer scheme as described in \cref{section:radiative_transfer}, accounting for scattering due to H$_2$ and He. We use the same molecule cross sections and quantum yields as described in \citet{Venot2012}.

We benchmark our chemical kinetics scheme with that of \citet{Venot2012} in \cref{section:chemical_kinetics_test}.

\subsection{Fully-Consistent Modelling}

Since the chemical kinetics scheme is consistently coupled with the radiative-convective scheme it allows the non-equilibrium chemical abundances to be solved consistently with the $P$--$T$ profile. Previous chemical kinetics models of hot Jupiter atmospheres supply the $P$--$T$ profile as a fixed model input, on which the chemical abundances are calculated \citep[e.g.][]{Line2010,Moses2011,Venot2012,Kopparapu2012}. Our approach allows changes in opacity, due to non-equilibrium chemistry, to feedback into the radiative-convective balance and hence maintain a consistent temperature structure.

Initially, we solve for the $P$--$T$ profile assuming local chemical equilibrium. That is, we solve for radiative-convective balance using the Gibbs energy minimisation scheme to re-calculate the abundances on every iteration of the Newton-Raphson solver. Once radiative-convective balance (\ref{equation:energy}) is found, a 1D $P$--$T$ profile and corresponding abundance profiles of all the chemical species is obtained.

To find the $P$--$T$ profile consistent with non-equilibrium chemistry the model is restarted, switching to the chemical kinetics scheme including vertical mixing and photochemistry. The chemical equilibrium abundances and corresponding $P$--$T$ profile are used to initialise the model. Note that in the absence of vertical mixing and photochemistry the derived abundances from the Gibbs minimisation scheme and the chemical kinetics scheme agree very well; see \cref{section:chemical_kinetics_test} and \citet{Venot2012}. As the continuity equation (\cref{equation:continuity}) is solved, as outlined above, we periodically solve for the new radiative-convective balance using the up-to-date chemical abundances which may have changed due to vertical mixing and/or photochemical processes. We re-converge to find radiative-convective balance every 10 timesteps of the DLSODES solver; this value was chosen as the chemical abundances do not vary by a large amount within 10 iterations.

For convergence of the consistent model we use the criteria that the error in the energy flux balance equation (\ref{equation:energy}) is $< 10^{-4}$. For the kinetics calculations we ensure that the maximum relative error of all the chemical species (max($dn_i/n_i$)) is $< 10^{-4}$ {\it and} the model has reached an integration time of 10$^{12}$ s. Overall model convergence is deemed to be reached when both of these criteria are satisifed simultaneously. We tested using a higher tolerance and found that this is a good balance between accuracy and performance. The resulting $P$--$T$ profile and non-equilibrium chemical abundance profiles are fully consistent with each other. In addition, we also perform a chemical kinetics calculation with the $P$--$T$ profile held fixed.

\subsection{Test Case Planets}

In this first application of \texttt{ATMO} to highly irradiated exoplanet atmospheres we choose to model two hot Jupiters; HD~189733b and HD~209458b. These atmospheres are well studied both observationally and theoretically. In this work we do not attempt to match the various observations of these atmospheres. Instead we focus on the theoretical implications of non-equilibrium chemistry on both the temperature structure and the emission spectrum.  

We model the pressure range 10$^3$ to 10$^{-5}$ bar with 100 vertical levels; we have tested with both higher and lower vertical resolutions and found that this a good balance between model accuracy (most importantly accurately capturing the location of the quenching point) and computational efficiency. We do not model lower pressures since the atmosphere transitions into the thermosphere around this region and a different modelling approach is required \citep[e.g.][]{Yelle2004,Koskinen2013}, requiring consideration of non-local thermodynamic equilibrium. Therefore, due to the lower pressure limit of our model, though we include photochemistry, we only capture the 'edge' of the photodissociation zone; photodissociations become the dominant driver of the chemistry for P $<$ 10$^{-5}$ bar. We assume an internal temperature of 100 K for both planets. For both cases we have assumed a solar composition with elemental abundances of \citet{Caffau2011}.

The planetary, orbital and stellar paramaters of \citet{Southworth2010} are adopted in both cases and are summarised in \cref{table:planet_parameters}. We use the Kurucz stellar spectra\footnote{http://kurucz.harvard.edu/stars.html} of both HD~189733 and HD~209458 to compute the short-wave irradiation. For HD 209458 we use an XUV spectrum based on the solar spectrum, as in \citet{Venot2012}, where the solar XUV spectrum is scaled to the temperature and radius of HD 209458 with a scaling factor of ($R_{\rm star}$/$R_{\rm o}$)$^2$($T_{\rm star}$/$T_{\rm o}$)$^4$ $\approx$ 1.8. For the HD~189733 XUV flux we follow the same approach as \citet{Moses2011}: the UV flux of a K2 V star, epsilon Eridani, taken from the CoolCAT database\footnote{http://casa.colorado.edu/~ayres/CoolCAT/}, was used in range 115-230 nm, the solar UV flux for wavelengths below this range, and the solar flux divided by ten for longer wavelengths.

\begin{table}
\caption{Parameters used in this study.} 
\centering 
\label{table:planet_parameters}
\begin{tabular}{c c c} 
\hline\hline 
&HD~209458b & HD~189733b \\
\hline 
Mass $M_J$  & 0.714 & 1.150 \\ 
Radius $R_J$ & 1.380 & 1.151 \\
Orbital Dist. AU & 0.047 & 0.031\\
Stellar Spectral Type & G0V & K1-K2 \\
Stellar Radius $R_{\rm Sun}$ & 1.162 & 0.752 \\ 
\hline
\end{tabular}
\end{table}

\section{Results}
\label{section:results}

In this section we present the $P$--$T$ profiles, chemical abundances and simulated spectra derived from our 1D atmosphere code \texttt{ATMO} for the two test case hot Jupiter atmospheres outlined in the previous section.

For both planets we present a series of models with different approaches to the calculation of the chemical abundances. We show models which are consistent with local chemical equilibrium (EQ). We also include non-consistent non-equilibrium (NEQ) models, with chemical abundances derived from chemical kinetics calculations, including the processes of vertical mixing and photochemistry, where we hold the $P$--$T$ profile fixed. Finally, we include consistent non-equilibrium (CNEQ) models where the abundances are, again, derived from chemical kinetics but the $P$--$T$ profile is not fixed and is instead allowed to vary to maintain radiative-convective balance.


\subsection{HD~189733b}

\begin{figure} 
	\centering
	\resizebox{\hsize}{!}{\includegraphics{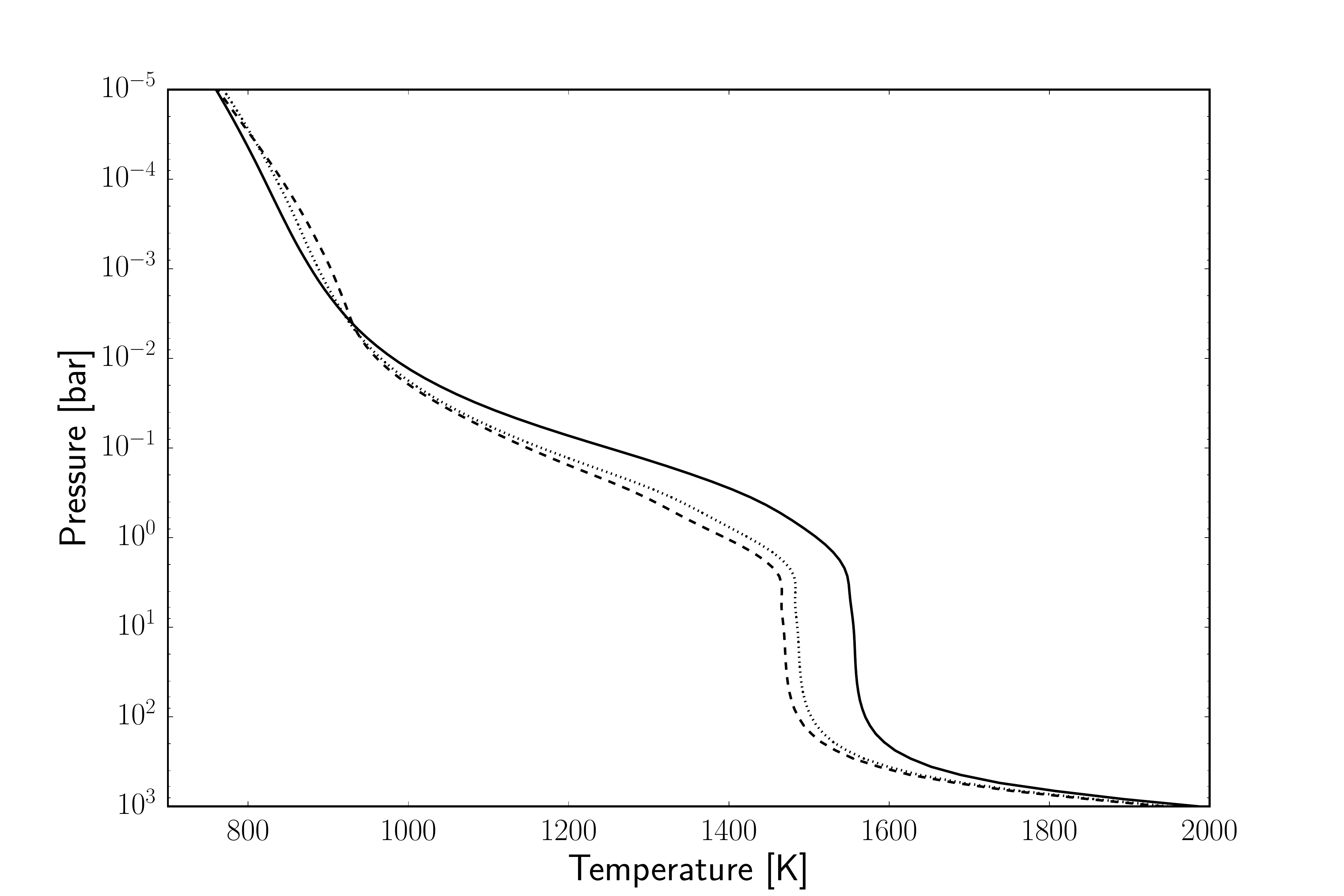}}
	\caption{The dayside average $P$--$T$ profiles for HD~189733b assuming equilibrium chemistry (EQ model, dashed), consistent non-equilibrium chemistry with $K_{zz}$ = 10$^{11}$ cm$^2$s$^{-1}$  (CNEQ model, solid) and with $K_{zz}$ = 10$^{9}$ cm$^2$s$^{-1}$ (CNEQ model, dotted). Note that the NEQ models referred to in the text use the same EQ $P$--$T$ profile as plotted in this figure.}
	\label{figure:pt_hd189} 
\end{figure}

\cref{figure:pt_hd189} shows the $P$--$T$ profiles of our EQ and CNEQ HD~189733b models using two different values for the $K_{zz}$ parameter, as described in \cref{section:model_description}. For the model with stronger vertical mixing there is a significant influence of non-equilibrium chemistry on the $P$--$T$ profile. The CNEQ $P$--$T$ profile is  $\sim$100 K warmer than the EQ profile for P $>$ 0.1 bar. In the model with the smaller $K_{zz}$ parameter the CNEQ $P$--$T$ profile is warmer than the EQ profile by about 15 K.

\begin{figure}
	\centering
	\resizebox{\hsize}{!}{\includegraphics{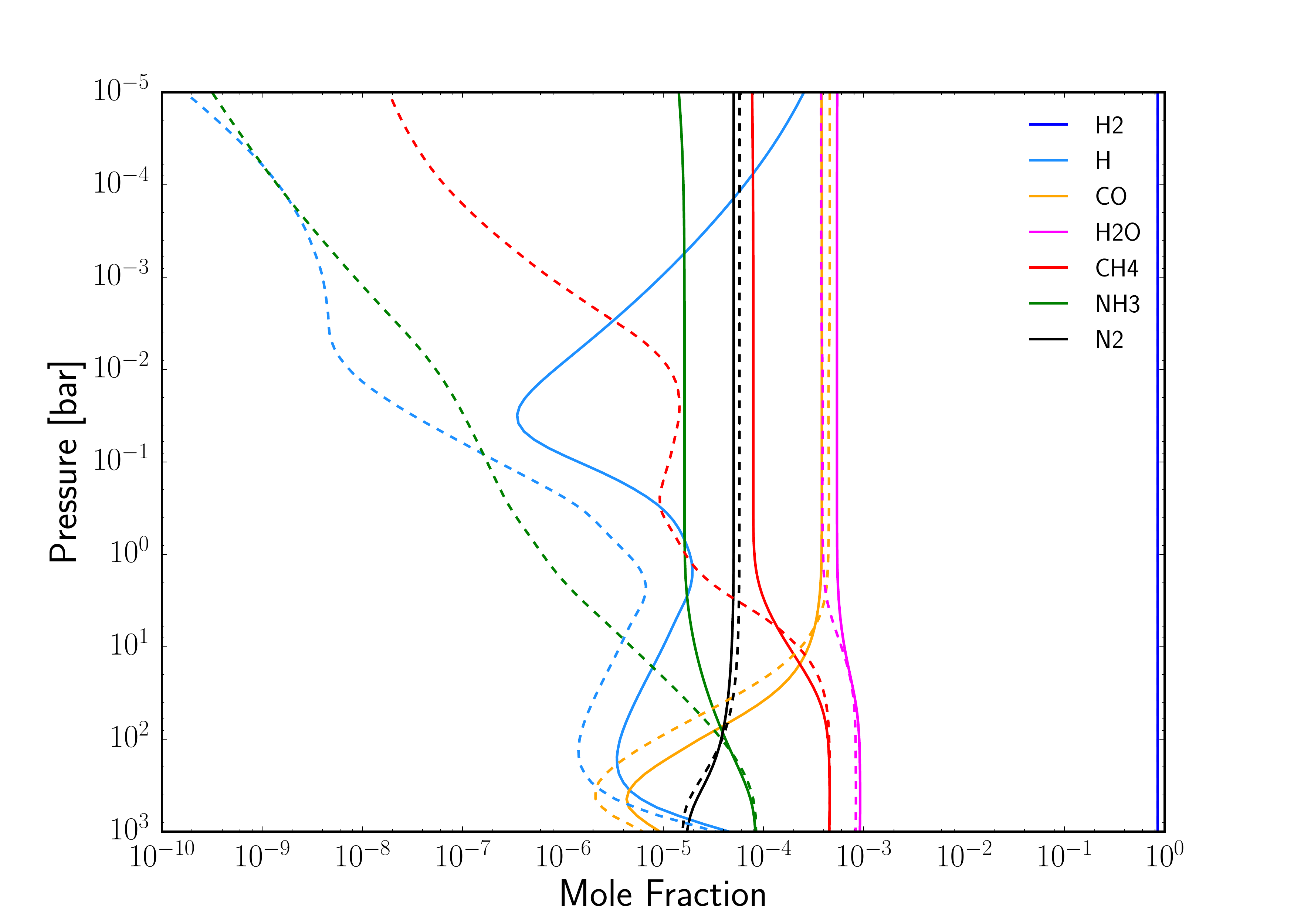}}
	\caption{The chemical abundances of the major chemical species for the HD~189733b model with abundances from the EQ calculation (dashed) and abundances from the CNEQ calculation including vertical mixing ($K_{zz}$ = 10$^{11}$ cm$^2$s$^{-1}$)  and photochemistry (solid).}
	\label{figure:chem_hd189_k11p} 
\end{figure}

\begin{figure}
	\centering
	\resizebox{\hsize}{!}{\includegraphics{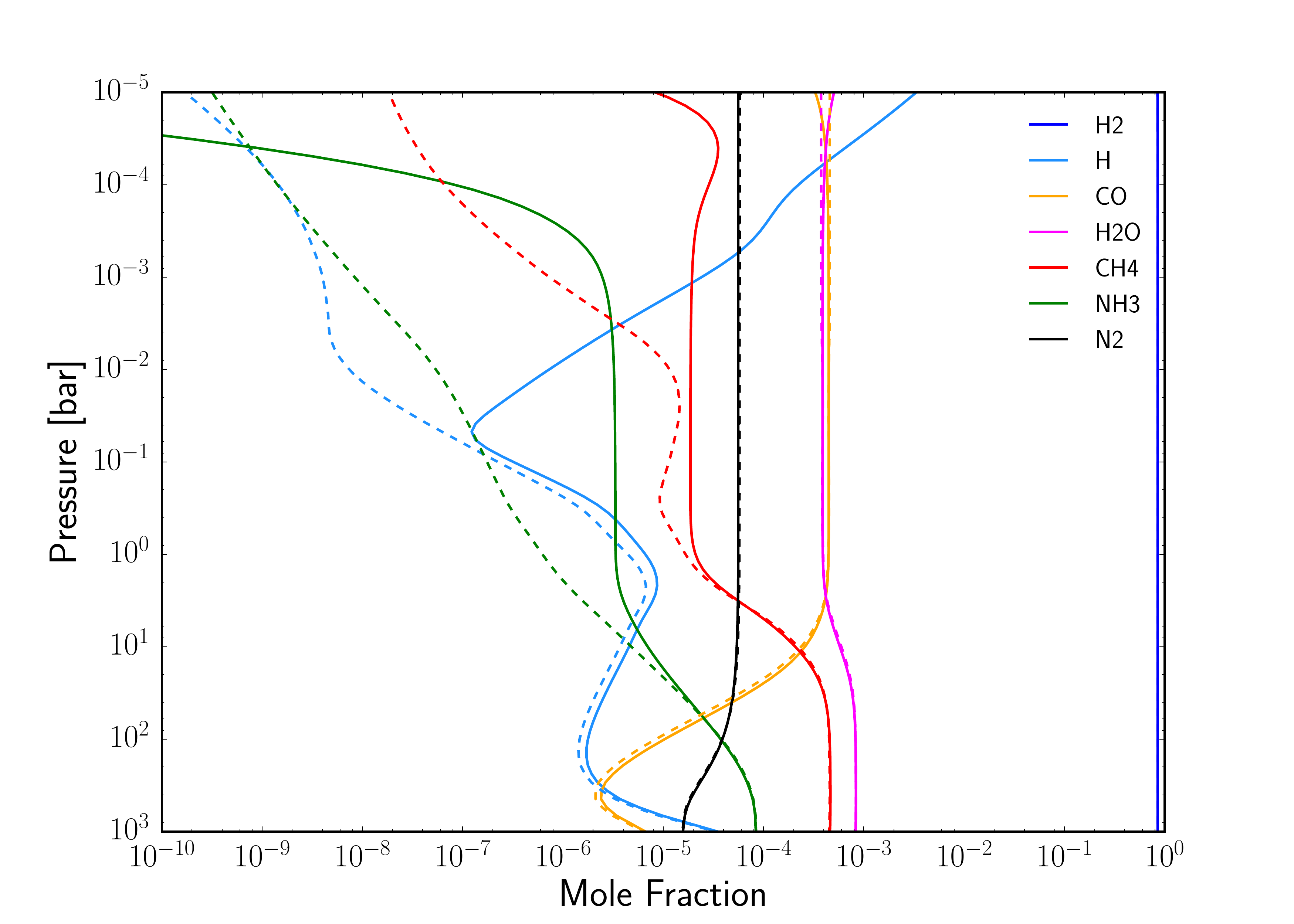}}
	\caption{Same as \cref{figure:chem_hd189_k11p} but using $K_{zz}$ = 10$^{9}$ cm$^2$s$^{-1}$.}
	\label{figure:chem_hd189_k9p} 
\end{figure}

The equilibrium and consistent non-equilibrium chemical abundances for these HD~189733b models are shown in \cref{figure:chem_hd189_k11p,figure:chem_hd189_k9p} for the models using $K_{zz}$ = 10$^{11}$ and 10$^{9}$ cm$^2$s$^{-1}$, respectively. Qualitatively, we find similar behaviour to previous studies \citep[][]{Moses2011,Venot2012} where the chemistry remains in chemical equilibrium in the hot deep atmosphere, the mid-regions of the atmosphere are dominated by vertical mixing, and photochemistry begins to become important for $P\sim$10$^{-5}$ bar.

The model using the stronger $K_{zz}$ shows significant increases in the abundance of both CH$_4$ and NH$_3$ compared to chemical equilibrium, which are quenched at around 10 bar and 100 bar, respectively. In chemical equilibrium, H$_2$O is more abundant than CO in the deep atmosphere, with CO becoming more abundant than H$_2$O at around 10 bar. This transition between the two molecules is removed with the inclusion of vertical mixing, as H$_2$O and CO are quenched below the transition, increasing the abundance of H$_2$O. The model with the smaller eddy diffusion coefficient shows a smaller increase in both CH$_4$ and NH$_3$ as their quench points are both shifted to lower pressures, reducing the quenched abundance. The effect on H$_2$O and CO is also much smaller in this model. 

\begin{figure}
	\centering
	\resizebox{\hsize}{!}{\includegraphics{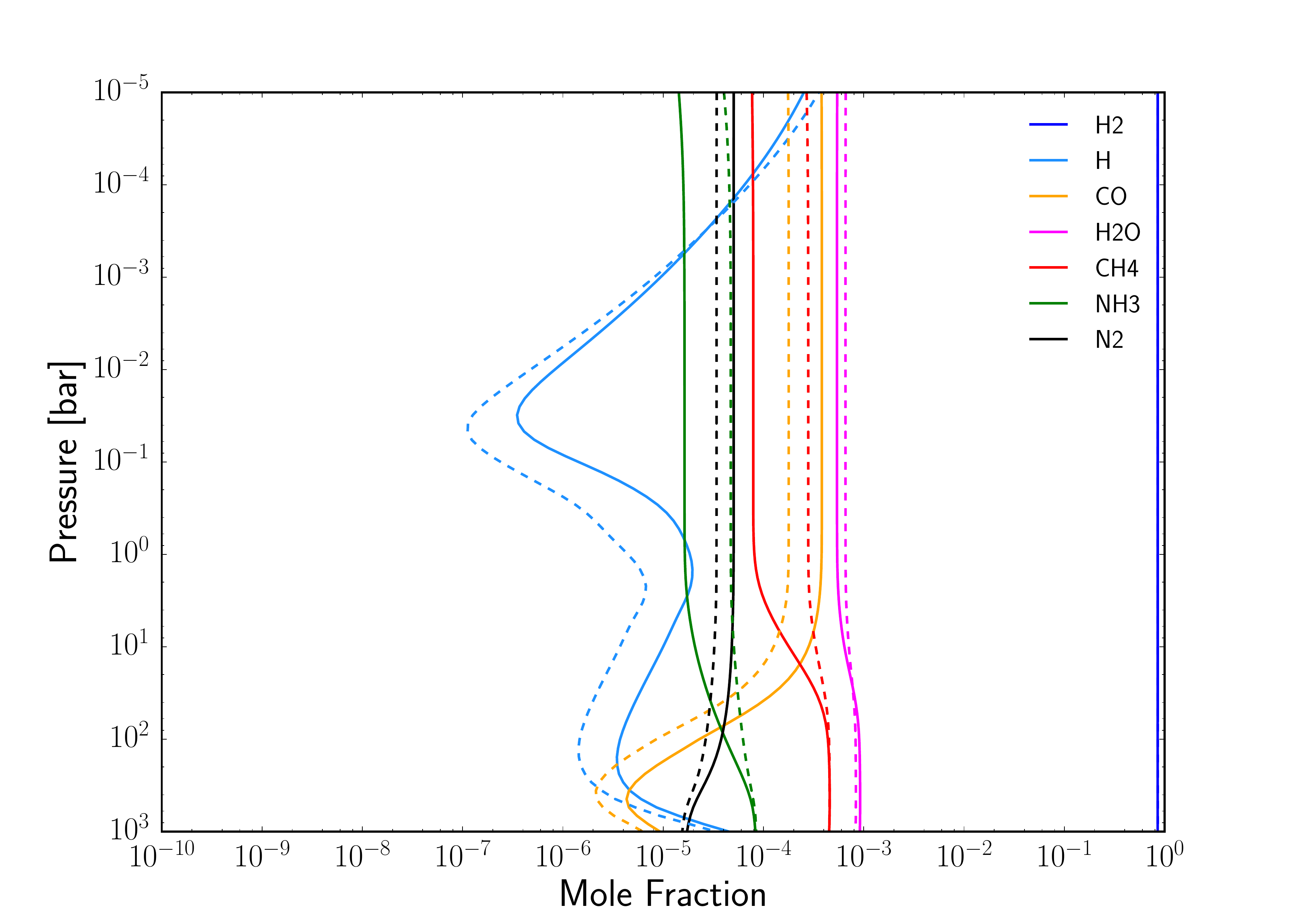}}
	\caption{A comparison of the chemical abundances between the CNEQ model (solid) and the NEQ model (dashed) for the HD~189733b $K_{zz}$ = 10$^{11}$ cm$^2$s$^{-1}$ case.}
	\label{figure:chem_hd189_k11p_pt_nopt} 
\end{figure}

The temperature change due to non-equilibrium chemistry has a feedback impact on the temperature-dependent chemical abundances.
Comparing the chemical abundances between the consistent and non-consistent (CNEQ and NEQ) models we see that for the strong vertical mixing case (\cref{figure:chem_hd189_k11p_pt_nopt}) the abundances of CO and CH$_4$ are essentially reversed, due to an increase in CO and a decrease in CH$_4$ in the CNEQ model. Similarly, in the NEQ model we find that NH$_3$ should be the dominant nitrogen species throughout the atmosphere, whereas in the CNEQ model we find that N$_2$ is the dominant nitrogen species (below 100 bar).

The temperature change induces different abundances via two processes. Firstly, the chemistry in the deep atmosphere remains in chemical equilibrium, and as the temperature increases the chemistry reaches a new chemical equilibrium with abundances which are consistent with the new temperature.

The second effect is caused by a shifting of the quench point. The quench point occurs at the pressure level where the chemical timescale $\tau_{\rm chem}$ is equal to the mixing timescale $\tau_{\rm mix}$. Since $\tau_{\rm chem}$ is dependent on temperature the quench point is shifted to a lower pressure level, in the case of a warmer atmosphere. This leads to a different quenched mole fraction effecting the quenched abundances for pressures below the quench point. This is complicated by the fact that the mole fractions of individual species have changed in the region of the quench point as they now exist in the new higher temperature chemical equilibrium, due to the first process explained.

\begin{figure}
	\centering
	\resizebox{\hsize}{!}{\includegraphics{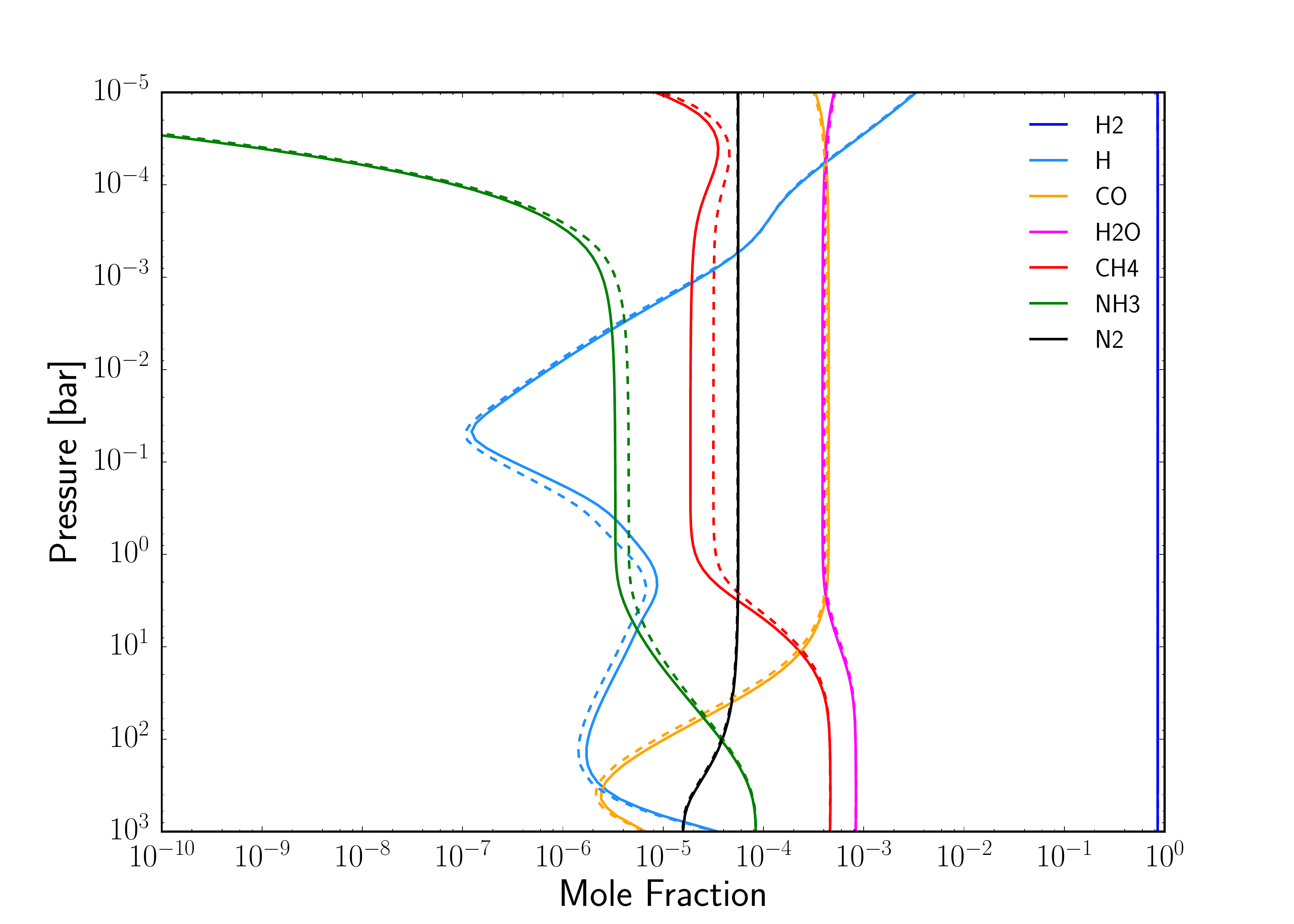}}
	\caption{Same as \cref{figure:chem_hd189_k11p_pt_nopt} but for the $K_{zz}$ = 10$^{9}$ cm$^2$s$^{-1}$ case.}
	\label{figure:chem_hd189_k9p_pt_nopt} 
\end{figure}

This process also occurs in the model with weaker vertical mixing (\cref{figure:chem_hd189_k9p_pt_nopt}) but to a smaller degree since the departure from chemical equilibrium and the induced temperature change is less.

Tests with only photochemistry included (i.e. without vertical mixing) show that the photochemistry has a negligible impact on the $P$--$T$ profile. Photodissociations become important for pressures below $\sim$10$^{-5}$ bar depending on the temperature and the UV flux. At these pressure levels the optical depth is small and changes in the chemical composition have negligible effect on the temperature structure. It is transport--induced quenching which is effective at higher pressures and higher optical depths which has the potential to alter temperature stucture. However, it may be that photochemical production of chemical species not included in our model, or not included as opacity sources in our model, could contribute to heating at low pressures; for example, ozone in the Earth atmosphere.

Overall, we find that for our model atmosphere of HD~189733b the process of transport-induced quenching causes temperature increases of up to 100 K between 10$^{-1}$-10$^{2}$ bar. This temperature increase, in turn, effects the calculated mole fractions by 1) inducing a new chemical equilibrium consistent with the higher temperature in the deep atmosphere and 2) shifting the quench point ($\tau_{\rm chem}$=$\tau_{\rm mix}$) to lower pressures and altering the quenched abundances at low pressures; these two processes act simultaneously. For the strong vertical mixing case presented here, this results in an atmosphere where CO is the dominant carbon species for P$<$10 bar. This contrasts with the non-consistent calculation where CH$_4$ is expected to be the dominant carbon species throughout the whole atmosphere. A similar process occurs for the N$_2$-NH$_3$ system leading to an N$_2$ dominated atmosphere in the CNEQ model but an NH$_3$ dominated atmosphere in the NEQ model.


\subsubsection{Simulated Emission Spectra}

\begin{figure}
	\centering
	\resizebox{\hsize}{!}{\includegraphics{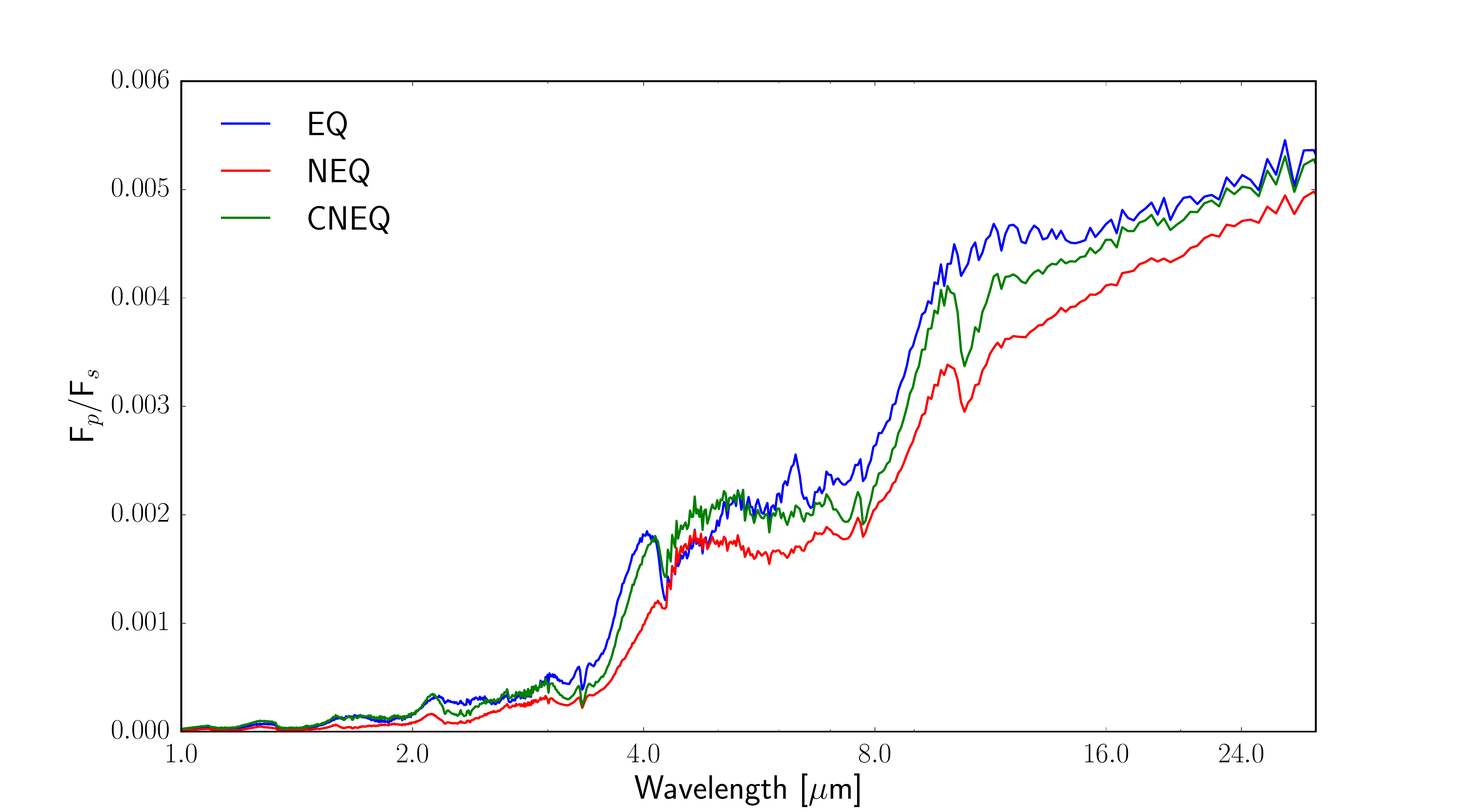}}
	\caption{The emission spectra of the HD~189733b model with $K_{zz}$ = 10$^{11}$ cm$^2$s$^{-1}$, showing calculations based on the EQ calculation (blue), the NEQ calculation (red) and the CNEQ calculation (green).}
	\label{figure:emiss_hd189_k11p} 
\end{figure}

In this section we present the simulated emission spectra for this series of models of HD~189733b. \cref{figure:emiss_hd189_k11p} shows the simulated emission spectrum for the stronger vertical mixing case for all three chemistry models (EQ, NEQ and CNEQ). Non-equilibrium chemistry has a strong impact on the simulated emission spectrum in the NEQ model. The NEQ spectrum has a significantly reduced flux ratio compared with the EQ model at almost all wavelengths. On the other hand, the CNEQ model shows a smaller discrepency with the EQ model, except at around 4.5 \textmu m where the CNEQ model shows a greater flux ratio than both the EQ and NEQ models.

\begin{figure}
	\centering
	\resizebox{\hsize}{!}{\includegraphics{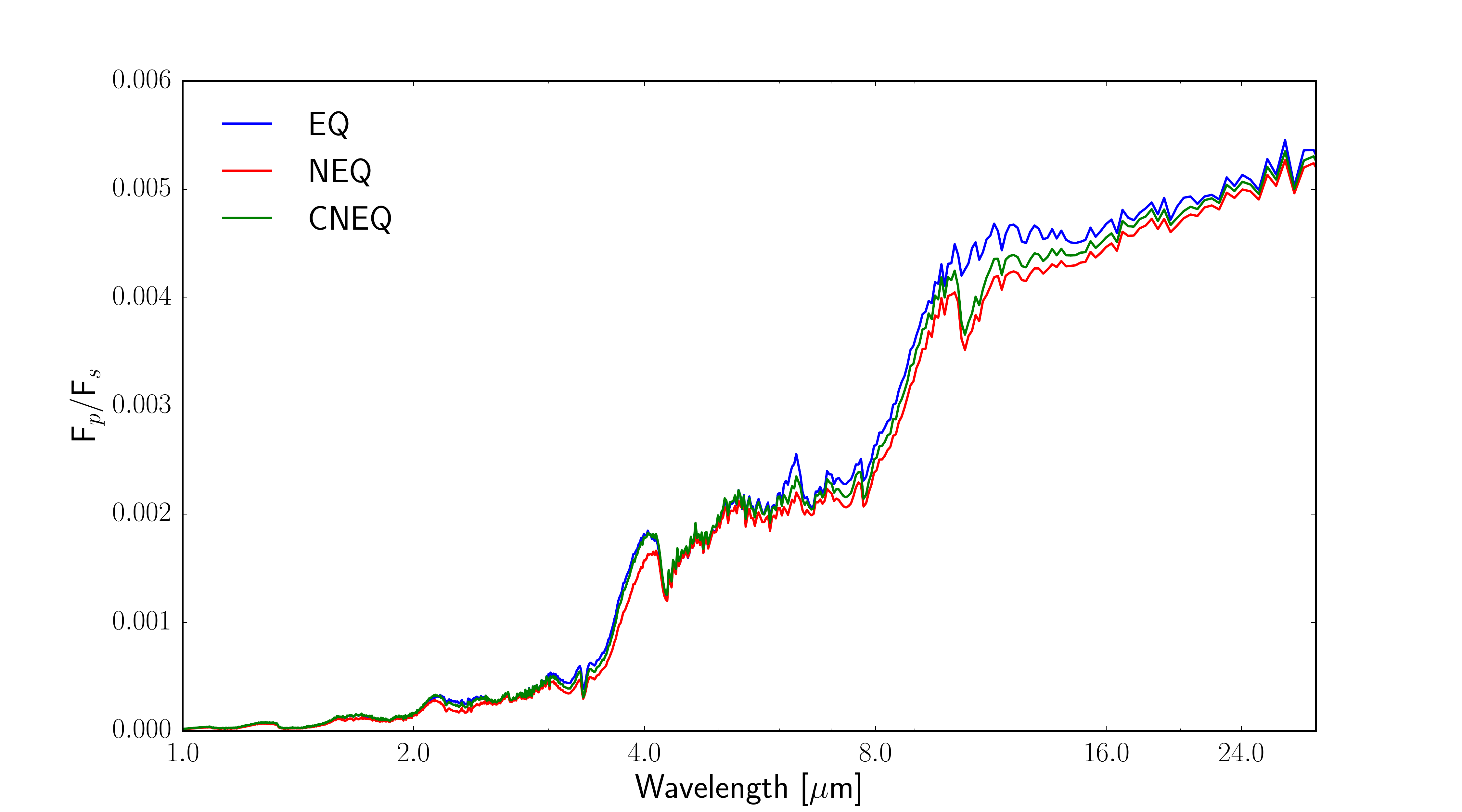}}
	\caption{Same as \cref{figure:emiss_hd189_k11p} but for the $K_{zz}$ = 10$^{9}$ cm$^2$s$^{-1}$ model.}
	\label{figure:emiss_hd189_k9p} 
\end{figure}

The model with the lower $K_{zz}$ value (\cref{figure:emiss_hd189_k9p}) shows similar trends, though the difference between the three chemistry cases is smaller as the departure from chemical equilibrium is not as strong. Our EQ and NEQ simulated spectra for this case agree well with the spectra of the 'thermochemical model' and 'photochemical model' of \citet[][their Fig. 11]{Moses2011}. We also find a reduction in flux at around 4 \textmu m and at longer wavelengths for our NEQ model. Interestingly, however, performing the chemical kinetics calculations consistently (CNEQ model) completely removes this signature of non-equilibrium chemistry at 4 \textmu m and also reduces the impact at longer wavelengths.

Most of the spectral features here are due to CH$_4$ (particularly around 3.6 \textmu m) and, at longer wavelengths, to NH$_3$ whilst CO is the dominant absorber around 4.5 \textmu m. Increases in the mole fractions of CH$_4$ and NH$_3$ due to transport-induced quenching increase the opacity in the wavelengths regions where they have absorption bands.


\subsection{HD~209458b}

We now present a series of models for the atmosphere of HD~209458b, which is warmer than the atmosphere of HD~189733b due to its orbit around a warmer G-type star; see \cref{table:planet_parameters}. There has been much debate about the presence of a thermal inversion in the atmosphere of HD~209458b, with early observations favouring the presence of an inversion \citep{Knutson2008,Beaulieu2010}. However, more recent re-analyses of these datasets suggest that this atmosphere does not contain a temperature inversion \citep{Diamond-Lowe2014,Evans2015}. In this study, we present both cases, as a way to explore a larger diversity of atmosphere types.

\begin{figure} 
	\centering
	\resizebox{\hsize}{!}{\includegraphics{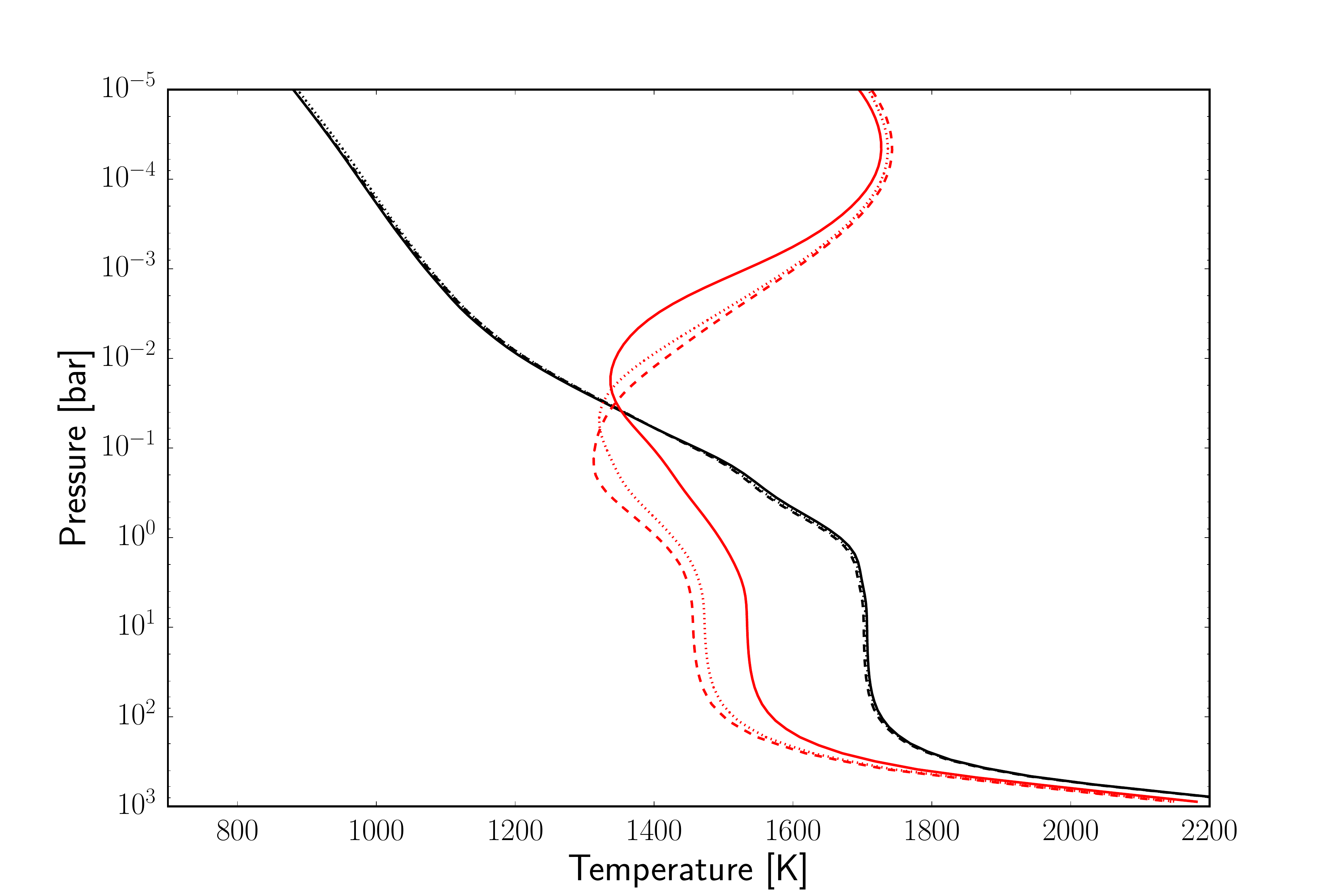}}
	\caption{The dayside average $P$--$T$ profiles for HD~209458b assuming equilibrium chemistry (EQ models, dashed), consistent non-equilibrium chemistry with $K_{zz}$ = 10$^{11}$ cm$^2$s$^{-1}$  (CNEQ models, solid) and with $K_{zz}$ = 10$^{9}$ cm$^2$s$^{-1}$ (CNEQ models, dotted), with the model including TiO and VO (red) and without TiO/VO (black). Note that the NEQ models referred to in the text use the same EQ $P$--$T$ profile as plotted in this figure.}
	\label{figure:pt_hd209} 
\end{figure}

\cref{figure:pt_hd209} shows the $P$--$T$ profiles for the two different atmosphere types of HD~209458b. In each case we show $P$--$T$ profiles for both the EQ and CNEQ chemistry models and with two different strenghs of vertical mixing; $K_{zz}$ = 10$^{11}$ and $K_{zz}$ = 10$^{9}$ cm$^2$s$^{-1}$, as in the previous section. For the rest of this analysis, however, we discuss only the $K_{zz}$ = 10$^{11}$ cm$^2$s$^{-1}$ case, where the impact of non-equilibrium chemistry is larger, and instead focus on the difference between the temperature inversion and non-temperature inversion cases.

For the case without a temperature inversion there are very small differences between the EQ and CNEQ $P$--$T$ profiles. For pressures greater than $\sim$0.1 bar the CNEQ model is marginally warmer than the EQ model; in the isothermal plateau region (1 bar $<$ $P$ $<$ 100 bar) the temperature is about $<$ 10 K warmer. On the other hand, the case with a temperature inversion shows a much greater discrepancy between the EQ and CNEQ models. The CNEQ $P$--$T$ profile is $>$ 100 K hotter than the EQ $P$--$T$ profile for pressures greater than 0.1 bar. At lower pressures the CNEQ $P$-$T$ profile is cooler than the EQ case by a similar amount. Interestingly, the position of the thermal inversion is also shifted to lower pressures.

\begin{figure}
	\centering
	\resizebox{\hsize}{!}{\includegraphics{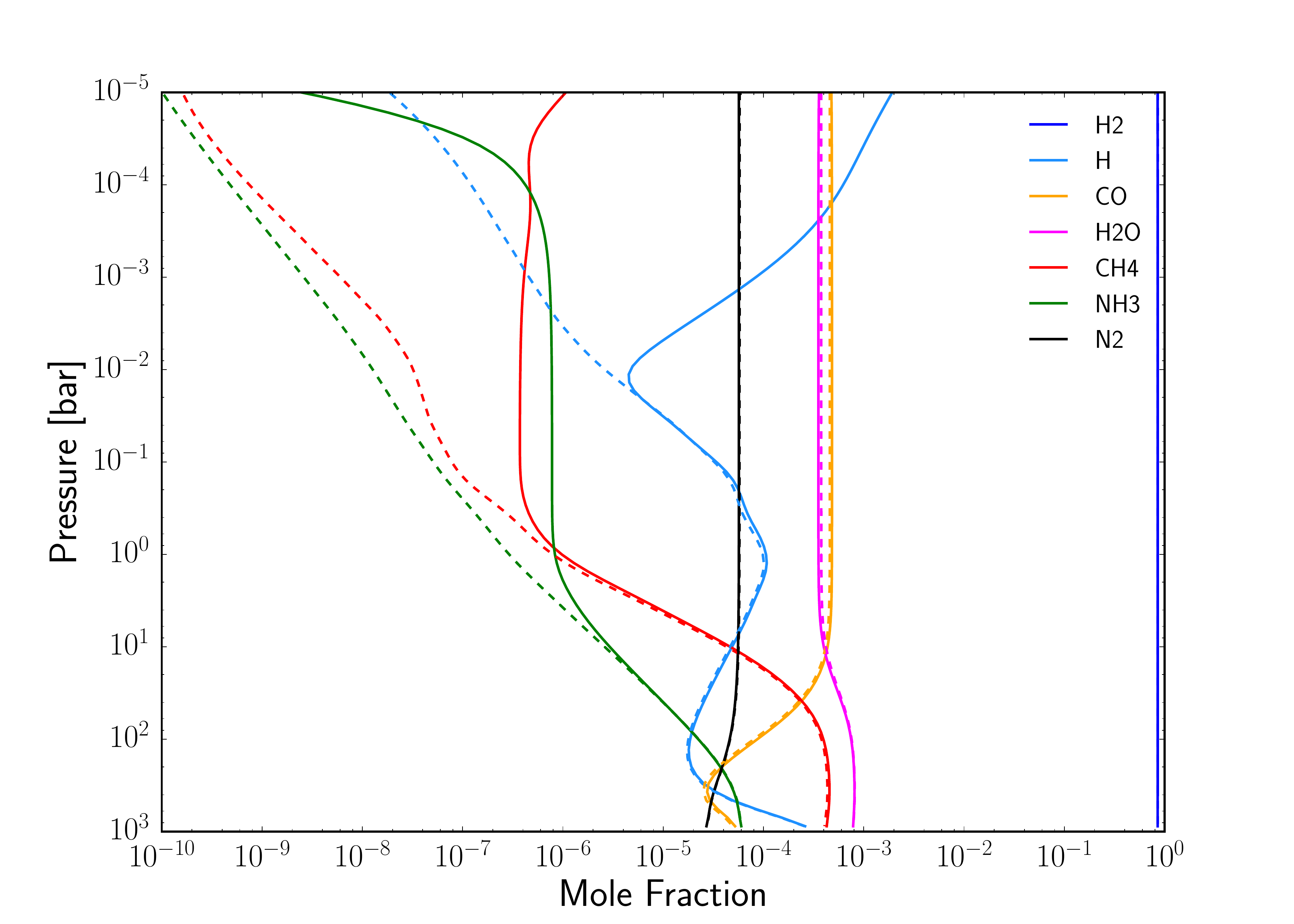}}
	\caption{The chemical abundances of the major chemical species for the HD~209458b model without a temperature inversion with abundances from the EQ calculation (dashed) and abundances from the CNEQ calculation including vertical mixing and photochemistry (solid); $K_{zz}$ = 10$^{11}$ cm$^2$s$^{-1}$.}
	\label{figure:chem_hd209_notiovo} 
\end{figure}

\begin{figure}
	\centering
	\resizebox{\hsize}{!}{\includegraphics{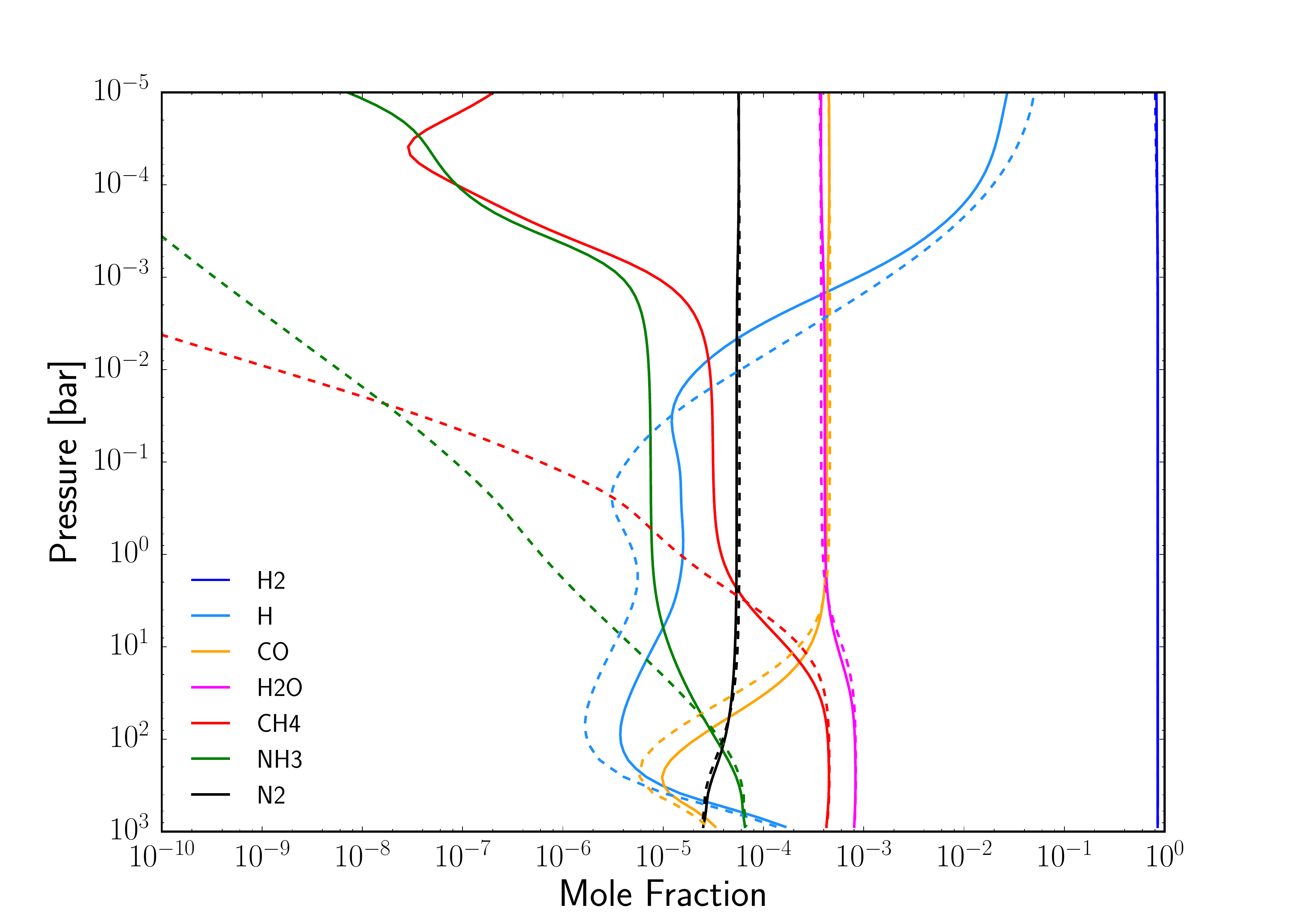}}
	\caption{Same as \cref{figure:chem_hd209_notiovo} but for the HD~209458b model with a temperature inversion.}
	\label{figure:chem_hd209_tiovo} 
\end{figure}

 \cref{figure:chem_hd209_notiovo} shows the abundances of the major chemical species for the model without a temperature inversion. NH$_3$ and CH$_4$ are quenched at around 10 and 1 bar respectively. This has the effect of increasing their abundances with respect to chemical equilibrium by several orders of magnitude for pressures lower than the quench point. Despite this, their molar fractions do not exceed 1$\times$10$^{-6}$ and N$_2$ and CO remain the dominant nitrogen and carbon species, which are unaffected by vertical mixing processes, and retain constant mixing ratios below 10 bar. The effects of photochemistry can be seen in the very upper regions of the model, particularly by the dissociation of NH$_3$ at 0.1 mbar and photochemical production of atomic H from 10 mbar.

\cref{figure:chem_hd209_tiovo} shows the abundances for the model with a temperature inversion which shows a much greater departure from chemical equilibrium. Though this model contains a hotter upper atmosphere, at depth the atmosphere is actually considerably cooler than the model without a temperature inversion. TiO and VO absorb visible photons at low pressures forming the temperature inversion. However, this reduces the flux of high energy photons which penetrate to depth and heat the lower atmosphere. This reduction of heating at high pressures leads to a cooler deep atmosphere. Due to the temperature dependence of the chemical timescale the quench point is shifted to higher pressures in the temperature inversion model, leading to larger quenched abundances of both CH$_4$ and NH$_3$. The quench points for CH$_4$ and NH$_3$ now lie at 10 and 100 bar, respectively. At around 1 mbar, where the temperature begins to increase again, the chemical timescale begins to speed up once more, and the species begin to move back towards their chemical equilibrium state, as seen in previous studies \citep{Moses2011,Venot2012}. 

The chemical abundances in the CNEQ and NEQ models (\cref{figure:chem_hd209_tiovo_k11p_pt_nopt}) show important differences. The NEQ model gives a CH$_4$ mole fraction around 5$\times$ larger than the CNEQ model. Similarly, the NH$_3$ mole fraction is $\sim$3.5$\times$ larger in the NEQ model compared with the CNEQ model. These discrepancies between the two models occur in the pressure range where observations are available. The differences are, again, due to the increase in temperature at depth which 1) changes the chemical equilibrium abundances in the deep atmosphere and 2) shifts the location of the quench point.

\begin{figure}
	\centering
	\resizebox{\hsize}{!}{\includegraphics{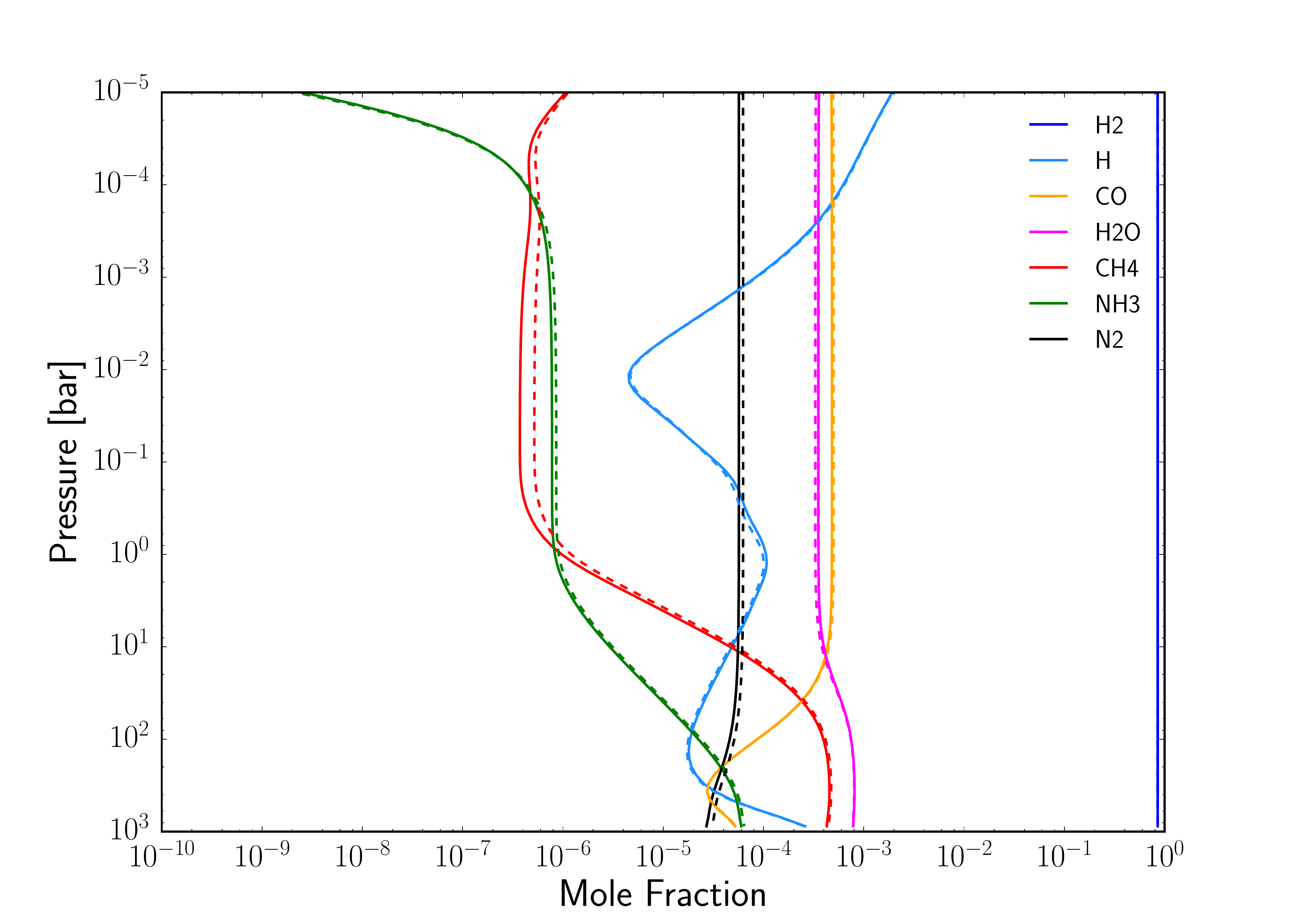}}
	\caption{A comparison of the chemical abundances between the CNEQ calculation (solid) and the NEQ calculation (dashed) for the HD~209458b model without a temperature inversion.}
	\label{figure:chem_hd209_notiovo_k11p_pt_nopt} 
\end{figure}

\begin{figure}
	\centering
	\resizebox{\hsize}{!}{\includegraphics{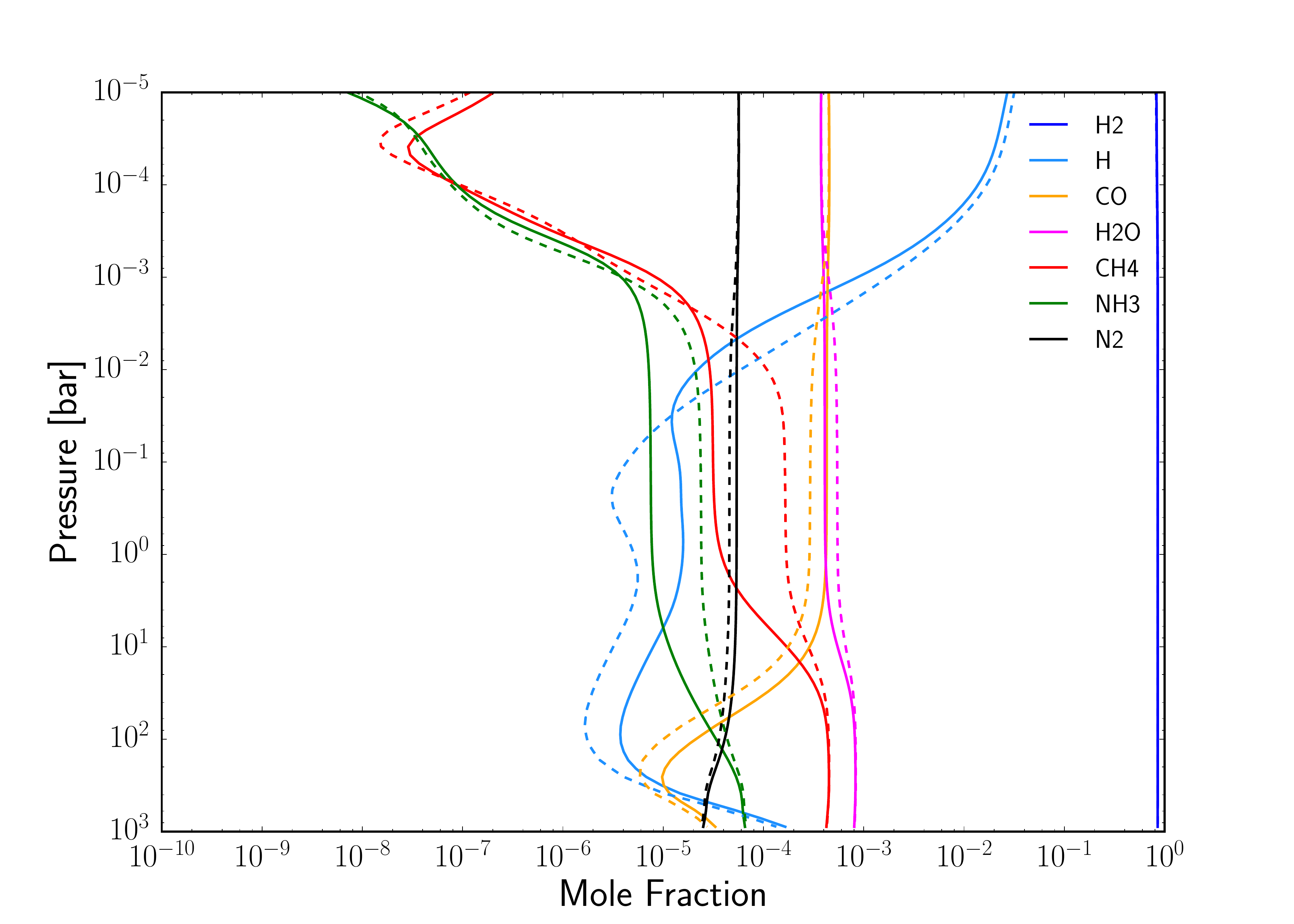}}
	\caption{Same as \cref{figure:chem_hd209_notiovo_k11p_pt_nopt} but for the HD~209458b model with a temperature inversion.}
	\label{figure:chem_hd209_tiovo_k11p_pt_nopt} 
\end{figure}

There is negligible difference between the CNEQ and NEQ abundances for the model without a temperature inversion (\cref{figure:chem_hd209_notiovo_k11p_pt_nopt}); only minor decreases in CH$_4$ and NH$_3$ in the CNEQ case. This is to be expected since the temperature difference between the CNEQ and NEQ model is also very small. 

We find that in the case of HD~209458b hosting a temperature inversion the deep atmosphere is cool enough to have significant non-equilibrium chemical abundances at depth. This leads to temperature increases of more than 100 K between 10$^{-1}$-10$^{2}$ bar, with decreases in the temperature at lower pressures. The location of the temperature inversion is also shifted to lower pressures. These temperature changes have important consequences on the chemical abundances and we find significantly smaller mole fractions of CH$_4$ and NH$_3$ in the consistent (CNEQ) model compared with the non-consistent (NEQ) model.

For the model without a temperature inversion the deep atmosphere is much hotter and the quench point exists at much lower pressures. This limits the influence of non-equilibrium chemistry and only a small temperature change is seen even for the strong vertical mixing case. Consequently there is little difference between the mole fractions of the consistent and non-consistent models.


\subsubsection{Simulated Emission Spectra}
\label{section:hd209_emission_spectra}

In this section we show the simulated emission spectra for the series of HD~209458b models presented above.

\begin{figure}
	\centering
	\resizebox{\hsize}{!}{\includegraphics{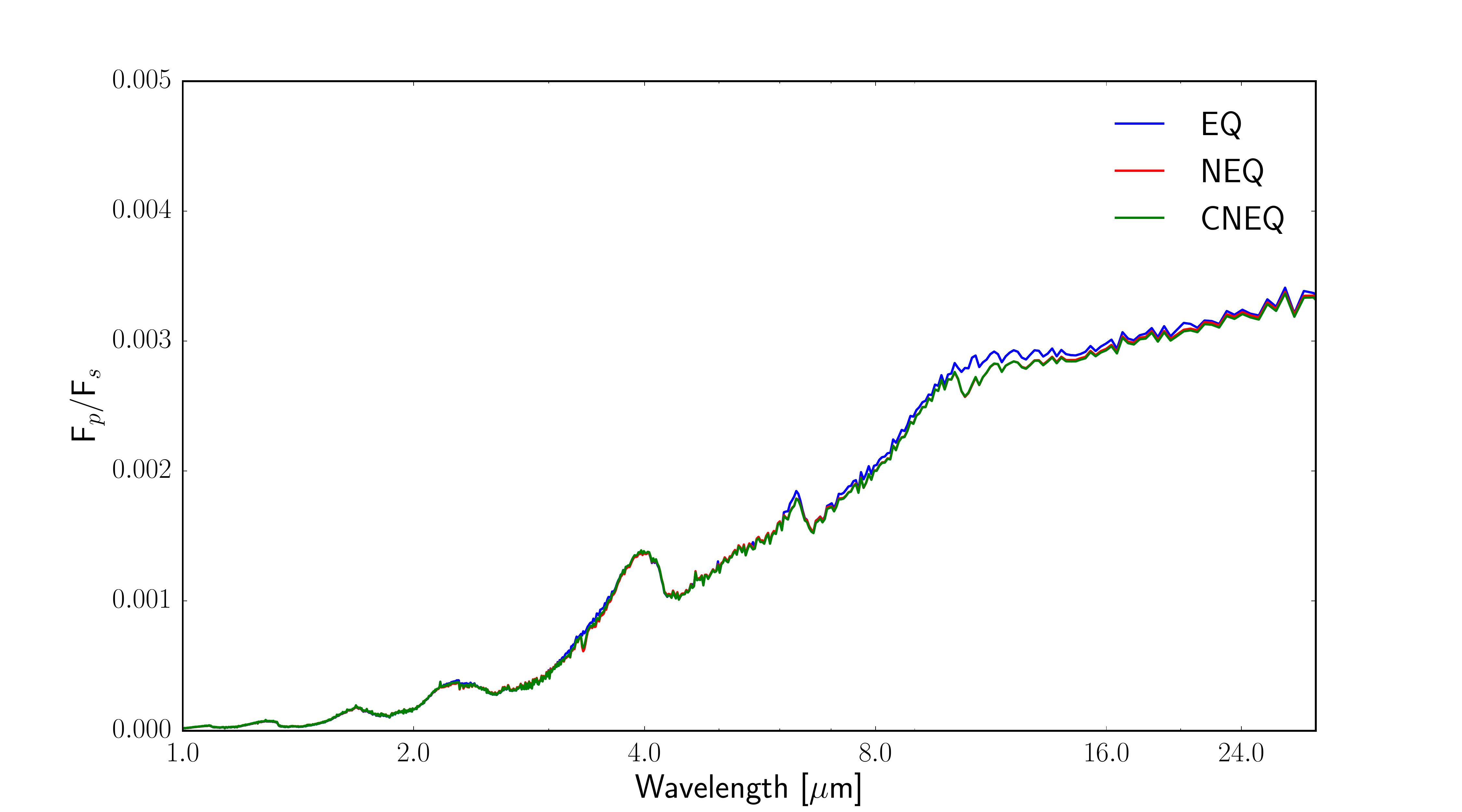}}
	\caption{The emission spectra of the HD~209458b model without a temperature inversion, showing calculations based on the EQ calculation (blue), the NEQ calculation (red) and the CNEQ calculation (green); $K_{zz}$ = 10$^{11}$ cm$^2$s$^{-1}$.}
	\label{figure:emiss_hd209_notiovo} 
\end{figure}

For the model without a temperature inversion (\cref{figure:emiss_hd209_notiovo}) there is a very small difference in the flux between the EQ, NEQ and CNEQ cases. This is not surprising since the departure from chemical equilibrium is small; and consequently the induced change in the $P$--$T$ profile is small.

\begin{figure}
	\centering
	\resizebox{\hsize}{!}{\includegraphics{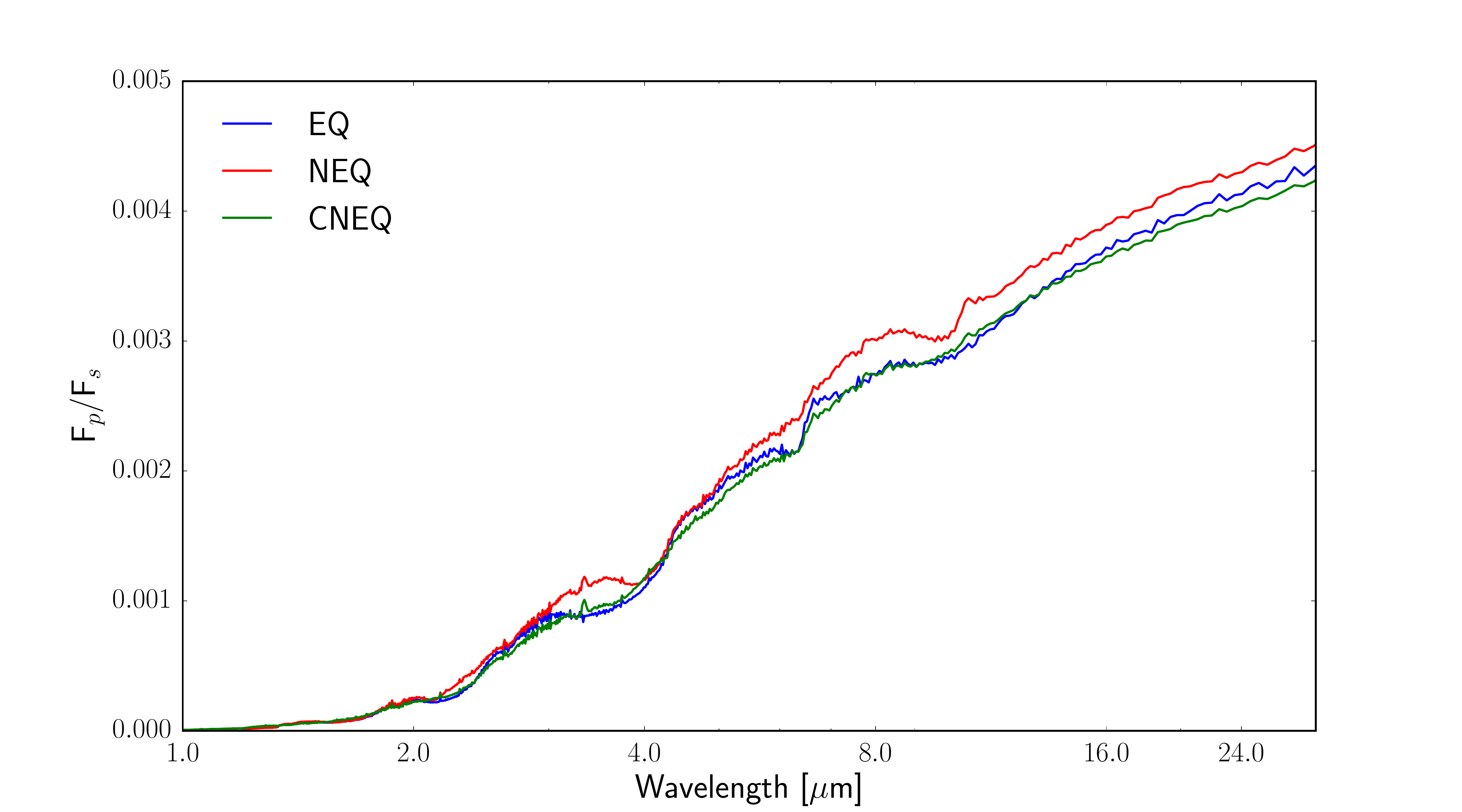}}
	\caption{Same as \cref{figure:emiss_hd209_notiovo} but for the HD~209458b model with a temperature inversion.}
	\label{figure:emiss_hd209_tiovo} 
\end{figure}

However, the models including a temperature inversion do show important differences (\cref{figure:emiss_hd209_tiovo}). The NEQ spectrum shows a greater flux over most of the wavelength range compared with the EQ and CNEQ spectra. In particular there is a large increase in flux around 3.6 \textmu m due to a large increase in the methane abundance due to transport--induced quenching. This is the opposite to what was found for the HD~189733b models where the flux was seen to decrease in the NEQ model. The primary difference between these models is that, in the HD~209458b model, at low pressures the temperature is increasing inversely with pressure, due to the presence of a temperature inversion. 

On the other hand, the EQ and CNEQ spectra are remarkably similar, despite the fact the abundances of methane and ammonia are driven far from chemical equilibrium and the $P$--$T$ profile is altered considerably. The increase in flux around 3.6 \textmu m and at longer wavelengths present in the NEQ model are not apparent in the CNEQ model, removing the signatures of non-equilibrium chemistry. The cause of this will be explained in detail in the next section.

In this particular case, the overall effect of calculating the non-equilibrium chemistry consistently with a coupled temperature structure, rather than on a fixed $P$--$T$ profile, is to reduce the influence of non-equilibrium chemistry on the emission spectrum, as found for the HD~189733b model, as the CNEQ spectrum tends back towards the EQ result.


\subsection{Energy Balance Considerations}

In this section we further investigate the differences resulting from a consistent treatment of calculating non-equilibrium chemical compositions.

\begin{figure}
	\centering
	\resizebox{\hsize}{!}{\includegraphics{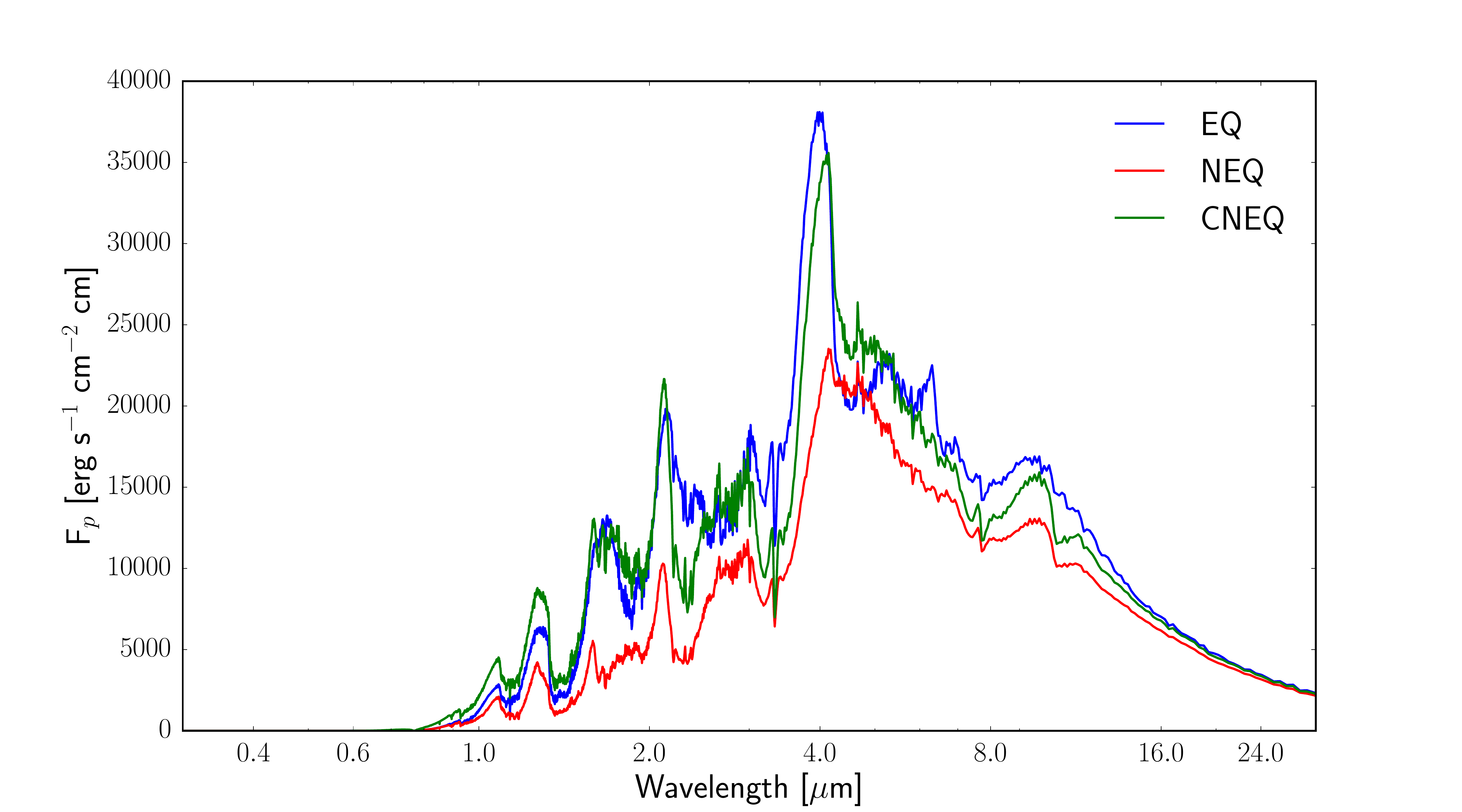}}
	\caption{The atmosphere emission spectrum of the HD~189733b model with $K_{zz}$ = 10$^{11}$ cm$^2$s$^{-1}$ showing calculations based on the EQ calculation (blue), the NEQ calculation (red) and the CNEQ calculation (green).}
	\label{figure:planet_emission_hd189_k11p} 
\end{figure}

\begin{figure}
	\centering
	\resizebox{\hsize}{!}{\includegraphics{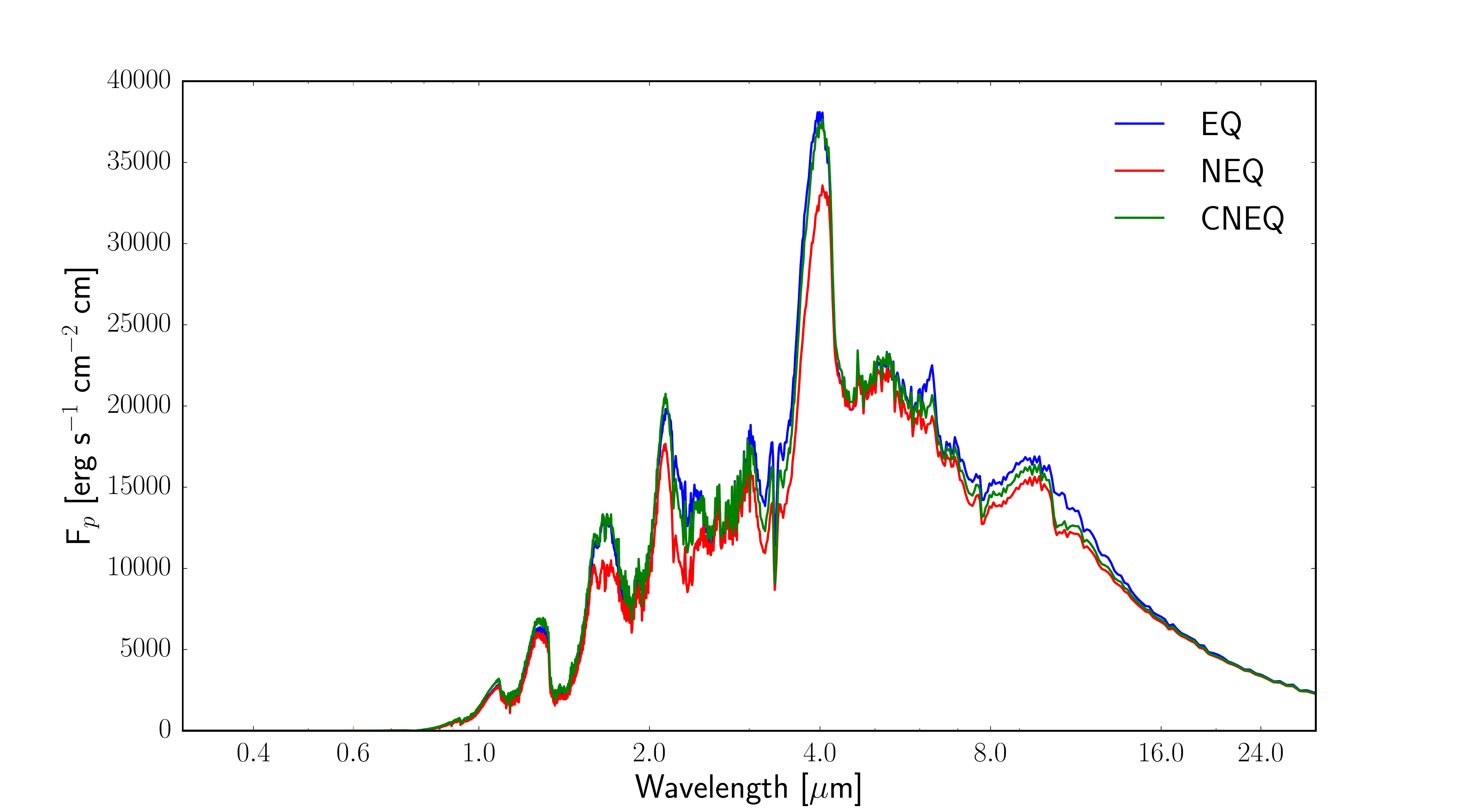}}
	\caption{Same as \cref{figure:planet_emission_hd189_k11p} but for the $K_{zz}$ = 10$^{9}$ cm$^2$s$^{-1}$ model.}
	\label{figure:emission_hd189_k9p} 
\end{figure}

\cref{figure:planet_emission_hd189_k11p,figure:emission_hd189_k9p} show the thermal emission spectra of the atmosphere (not divided by $F_{\rm star}$) for the HD~189733b models with $K_{zz}$ = 10$^{11}$ and 10$^{9}$ cm$^2$s$^{-1}$, respectively. In both cases, the emission is lower at all wavelengths for the NEQ models, compared to both the EQ and CNEQ models. This discrepancy is larger for the model with stronger vertical mixing. By eye, one can already see that the total amount of energy emission of the atmosphere (i.e. the wavelength-integrated flux) is less in the NEQ model, compared with the EQ and CNEQ models.


Indeed, this is the case, as shown in \cref{table:integrated_flux_hd189_k11p,table:integrated_flux_hd189_k9p} which present the integrated top of atmosphere flux (not including the reflected component) and corresponding blackbody temperatures for both $K_{zz}$ cases and for all chemistry models.  In both $K_{zz}$ cases, the integrated flux for the EQ and CNEQ models agree well with each other, conserving the total amount of energy being emitted by the atmosphere. On the other hand, the NEQ models show strongly reduced integrated fluxes. The integrated flux in the NEQ models is $\sim$38\% and $\sim$11\% smaller than the EQ integrated flux for the $K_{zz}$ = 10$^{11}$ cm$^2$s$^{-1}$ and $K_{zz}$ = 10$^{9}$ cm$^2$s$^{-1}$ models, respectively.

These calculations show that the NEQ models do not conserve energy and the model atmosphere is not in a state of energy balance. The incoming energy (irradiation and internal heating) has not changed, only the chemical abundance profiles have changed, yet the atmosphere is emitting less energy. However, the CNEQ models do conserve the amount of energy being lost by the atmosphere, as the integrated flux is equivalent to that found for the EQ models.

\begin{table}
\caption{Integrated flux and corresponding blackbody temperatures for the HD~189733b model with $K_{zz}$ = 10$^{11}$ cm$^2$s$^{-1}$.} 
\centering 
\label{table:integrated_flux_hd189_k11p}
\begin{tabular}{c c c }
\hline\hline 
&Flux (kWm$^{-2}$) & T$_{BB}$ (K) \\
\hline
EQ & 106.6 & 1171 \\
NEQ & 66.1 & 1039 \\
CNEQ & 106.4 & 1170\\
\hline
\end{tabular}
\end{table}

\begin{table}
\caption{Integrated flux and corresponding blackbody temperatures for the HD~189733b model with $K_{zz}$ = 10$^{9}$ cm$^2$s$^{-1}$.} 
\centering 
\label{table:integrated_flux_hd189_k9p}
\begin{tabular}{c c c }
\hline\hline 
&Flux (kWm$^{-2}$) & T$_{BB}$ (K) \\
\hline
EQ & 106.6 & 1171 \\
NEQ & 94.6 & 1137 \\
CNEQ & 106.6 & 1171\\
\hline
\end{tabular}
\end{table}

\begin{figure}
	\centering
	\resizebox{\hsize}{!}{\includegraphics{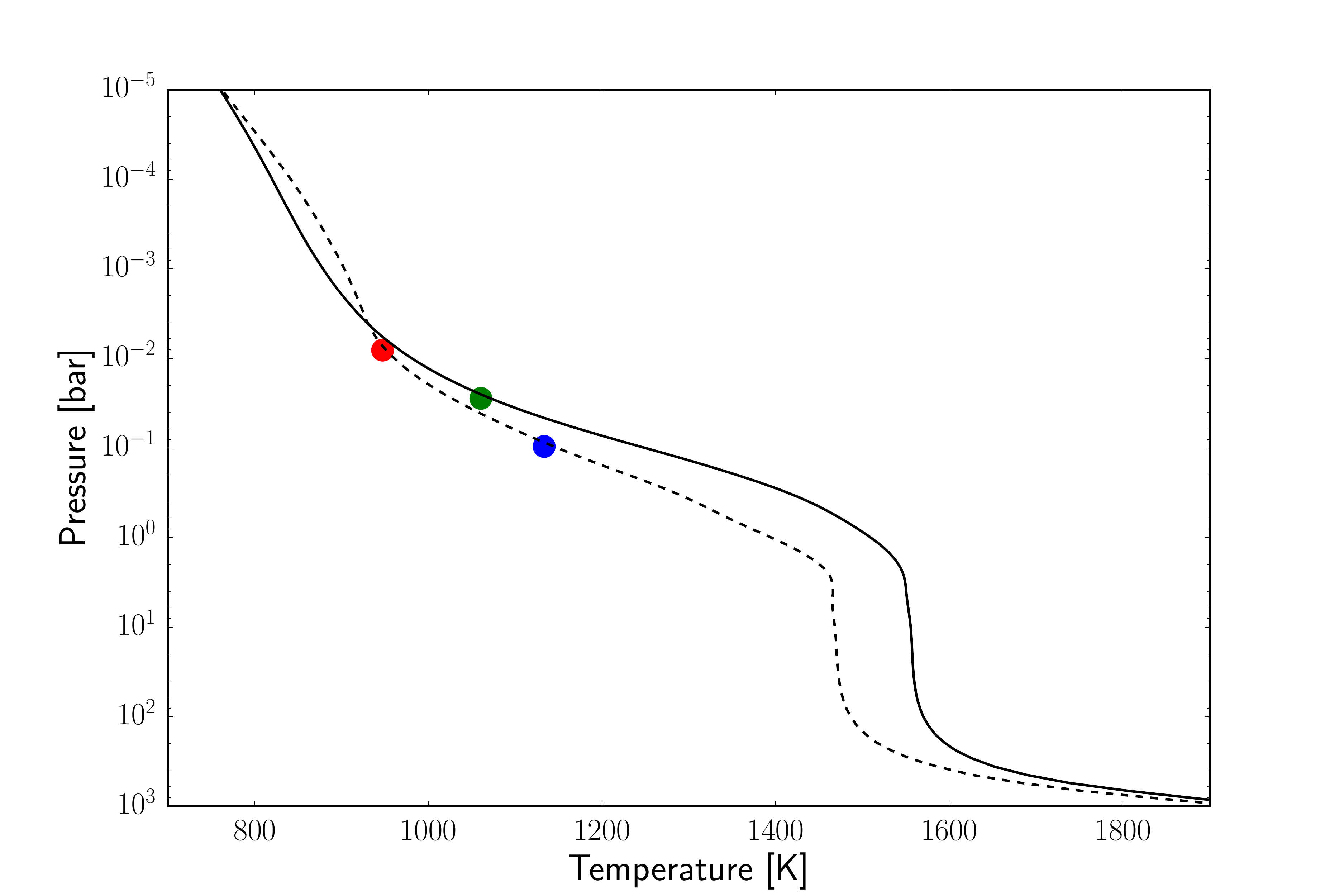}}
	\caption{The location of the peak emission in the 3.6 \textmu m Spitzer band for the HD~189733b $K_{zz}$ = 10$^{11}$ cm$^2$s$^{-1}$ model with the EQ calculation (blue), NEQ calculation (red) and CNEQ calculation (green).}
	\label{figure:3.6micron_hd189_k11p} 
\end{figure}

\begin{figure}
	\centering
	\resizebox{\hsize}{!}{\includegraphics{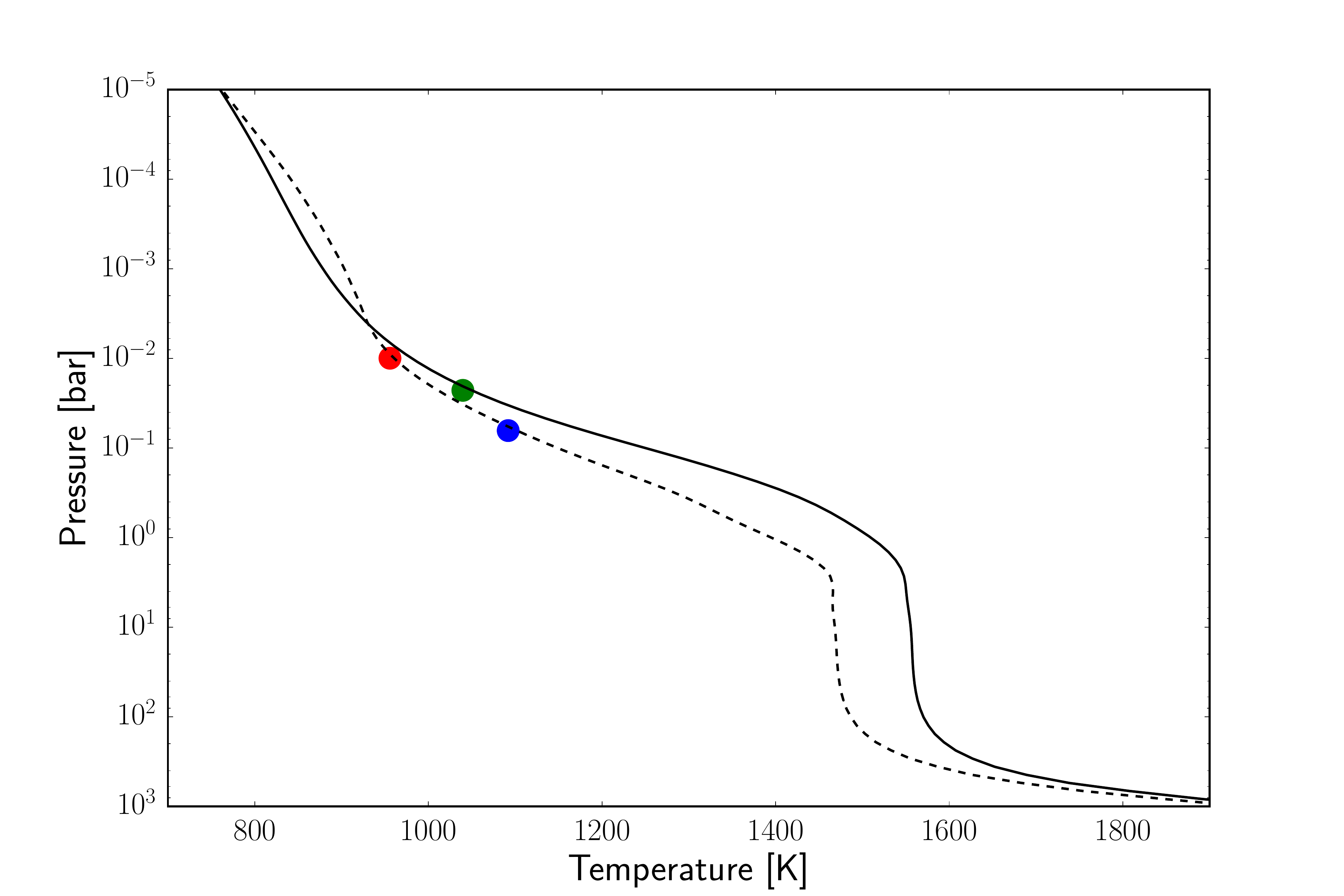}}
	\caption{As \cref{figure:3.6micron_hd189_k11p} for the Spitzer/IRAC 8.0 \textmu m channel.}
	\label{figure:8micron_hd189_k11p} 
\end{figure}

To understand this further, we show the pressure level of peak emission (i.e. the photosphere) in  \cref{figure:3.6micron_hd189_k11p,figure:8micron_hd189_k11p} for the 3.6 \textmu m and 8.0 \textmu m Spitzer/IRAC channels, respectively. Here the pressure level of the photosphere is taken as the maximum of the contribution (or weighting) function \citep[e.g.][]{Knutson2009,Griffith1998}. In both cases, the pressure level of the photosphere is shifted to lower pressures, and lower temperatures, in the NEQ model. This is a result of increased opacity due transport-induced quenching of CH$_4$ and NH$_3$. 

Since the emission flux is strongly dependent on temperature, the shifting of the photosphere to lower temperatures results in a decreased emission in this wavelength band. Indeed, this occurs not just in this wavelength band but also at other points where CH$_4$/NH$_3$ absorb and is evident in the decreased integrated flux value previously shown for the NEQ models. For the CNEQ model, though the photosphere is shifted to lower pressures the temperature at this lower pressure is increased compared with the EQ/NEQ model. Here we see the $P$--$T$ profile adapting to maintain radiative-convective equilibrium and energy balance in reaction to the changing chemical composition due to non-equilibrium chemistry.


\begin{figure}
	\centering
	\resizebox{\hsize}{!}{\includegraphics{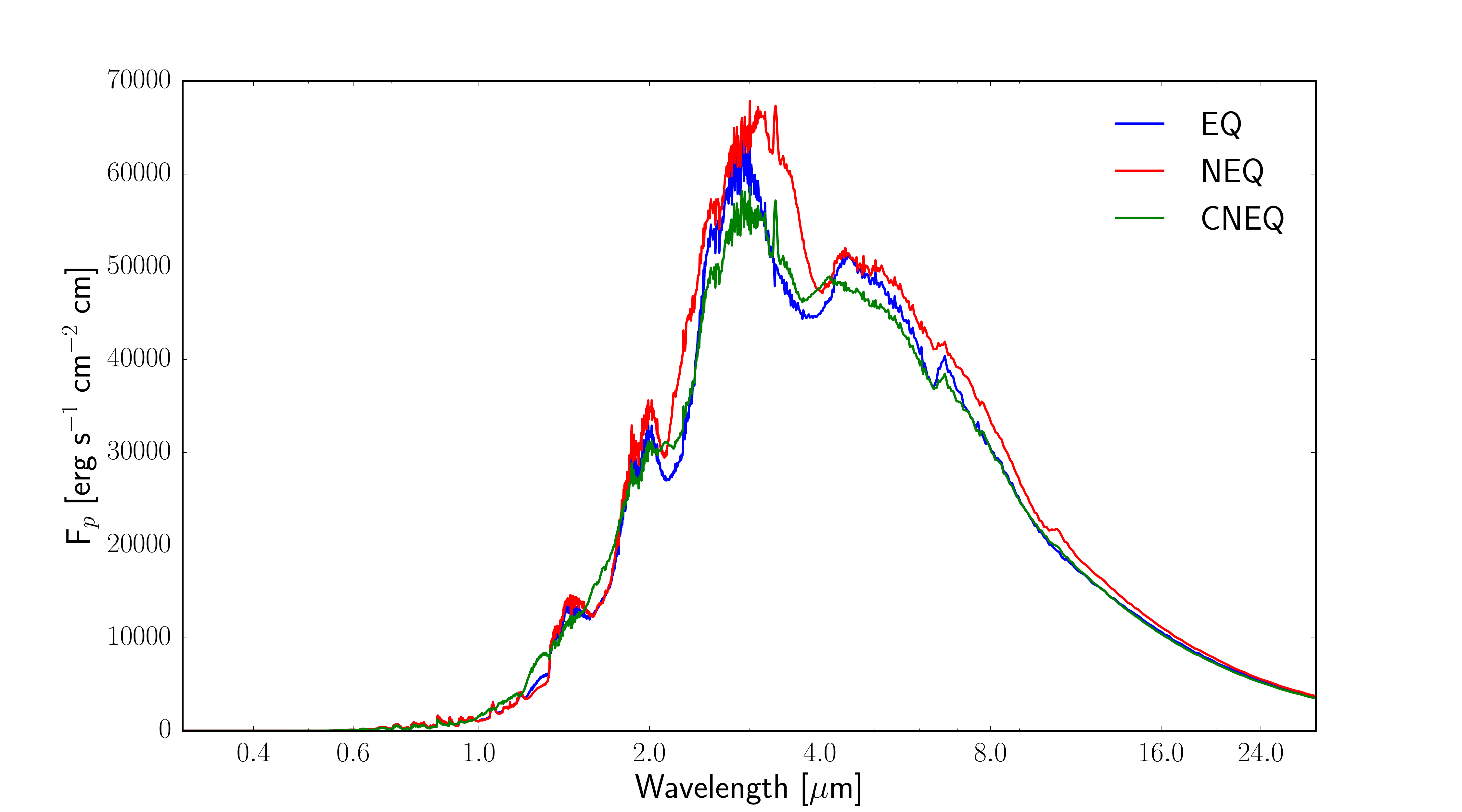}}
	\caption{The atmosphere emission spectrum of the HD~209458b model with a temperature inversion showing calculations based on the EQ calculation (blue), the NEQ calculation (red) and the CNEQ calculation (green); $K_{zz}$ = 10$^{11}$ cm$^2$s$^{-1}$.}
	\label{figure:planet_emission_hd209_tiovo} 
\end{figure}

\begin{figure}
	\centering
	\resizebox{\hsize}{!}{\includegraphics{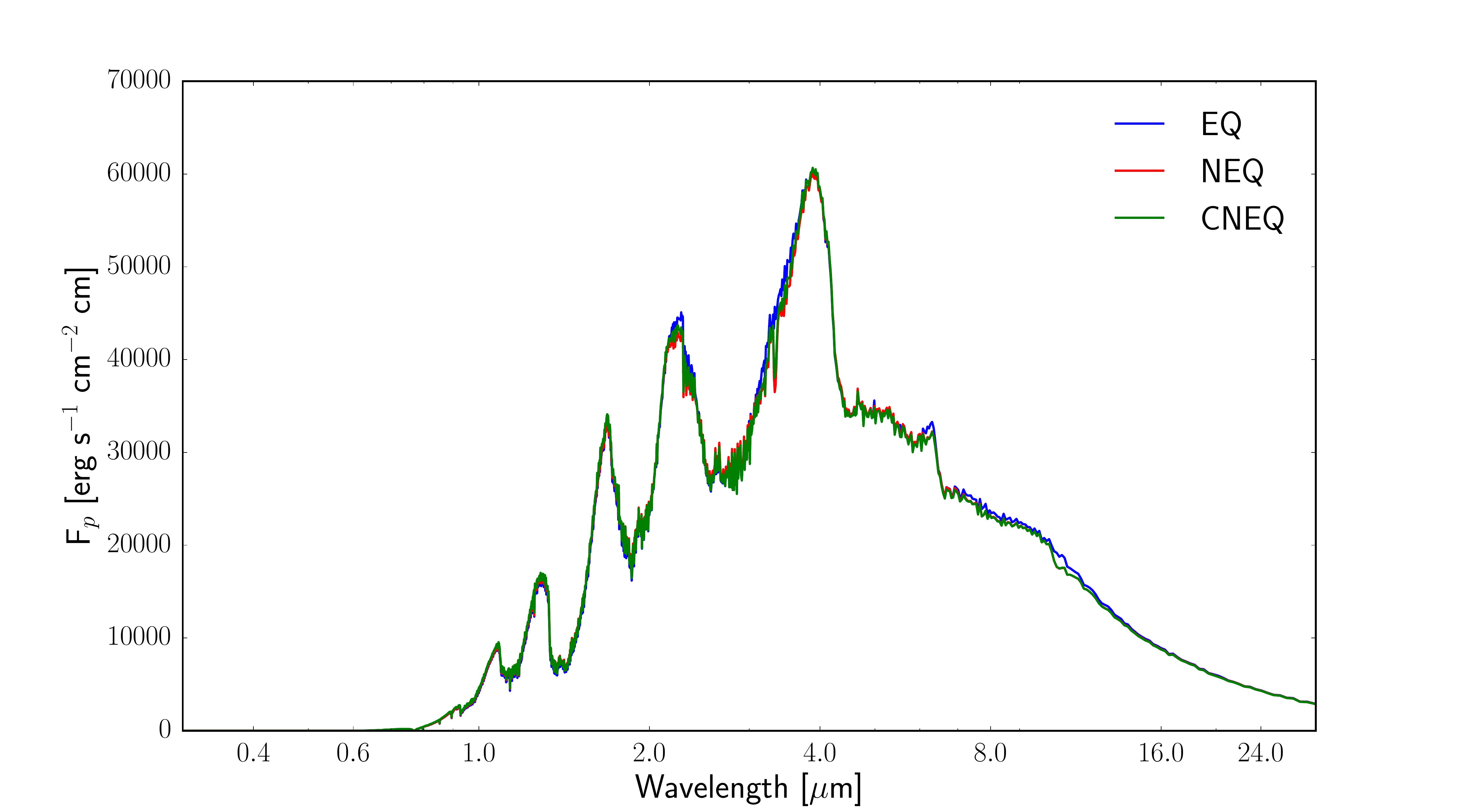}}
	\caption{Same as \cref{figure:planet_emission_hd209_tiovo} but for the model without a temperature inversion; $K_{zz}$ = 10$^{11}$ cm$^2$s$^{-1}$.}
	\label{figure:planet_emission_hd209_notiovo} 
\end{figure}

The story is very similar for the models of HD~209458b. \cref{figure:planet_emission_hd209_tiovo,figure:planet_emission_hd209_notiovo} show the top of atmosphere emission spectrum for the HD~209458b models with and without a temperature inversion, respectively, in each case for the EQ, NEQ and CNEQ models. The model without a temperature inversion shows a negligible difference between all three cases. In the temperature inversion model, the NEQ case shows a greater flux at all wavelengths compared with the EQ case. In particular, there is a large increase in flux between 3 and 4 \textmu m, which roughly corresponds to the wavelength of peak emission. Methane dominates the absorption around 3.6 \textmu m, which is increased in abundance by transport-induced quenching.

Again, by eye, it is possible to see that the NEQ models do not conserve the wavelength-integrated flux.
Indeed, \cref{table:integrated_flux_hd209_tiovo} shows the integrated flux of the atmosphere and the corresponding blackbody temperature for the temperature inversion models, where the integrated flux for the NEQ calculation is $\sim$10\% greater than the EQ model. On the other hand, the integrated flux for the CNEQ model is in excellent agreement with the EQ model. Likewise, \cref{table:integrated_flux_hd209_notiovo} shows the same information for the HD~209458b model without a temperature inversion. In this case, the discrepancy between EQ and NEQ is smaller. The EQ and CNEQ models show consistent outgoing fluxes, however, the NEQ model shows a very small $\sim$1\% reduction in integrated flux. 

For the model without a temperature inversion, the NEQ model shows a decreased integrated flux, similar to what was found for HD~189733b. However, the model with a temperature inversion shows an increased integrated flux for the NEQ model.


\begin{table}
\caption{Integrated flux and corresponding blackbody temperatures for the HD~209458b temperature inversion model; $K_{zz}$ = 10$^{11}$ cm$^2$s$^{-1}$.} 
\centering 
\label{table:integrated_flux_hd209_tiovo}
\begin{tabular}{c c c }
\hline\hline 
&Flux (kWm$^{-2}$) & T$_{BB}$ (K) \\
\hline
EQ & 235.4 & 1427 \\
NEQ & 256.5 & 1458 \\
CNEQ & 235.4 & 1428\\
\hline
\end{tabular}
\end{table}

\begin{table}
\caption{Integrated flux and corresponding blackbody temperatures for the HD~209458b model without temperature inversion, $K_{zz}$ = 10$^{11}$ cm$^2$s$^{-1}$.} 
\centering 
\label{table:integrated_flux_hd209_notiovo}
\begin{tabular}{c c c }
\hline\hline 
&Flux (kWm$^{-2}$) & T$_{BB}$ (K) \\
\hline
EQ & 222.2 & 1407 \\
NEQ & 221.9 & 1406 \\
CNEQ & 222.7 & 1408\\
\hline
\end{tabular}
\end{table}

\begin{figure}
	\centering
	\resizebox{\hsize}{!}{\includegraphics{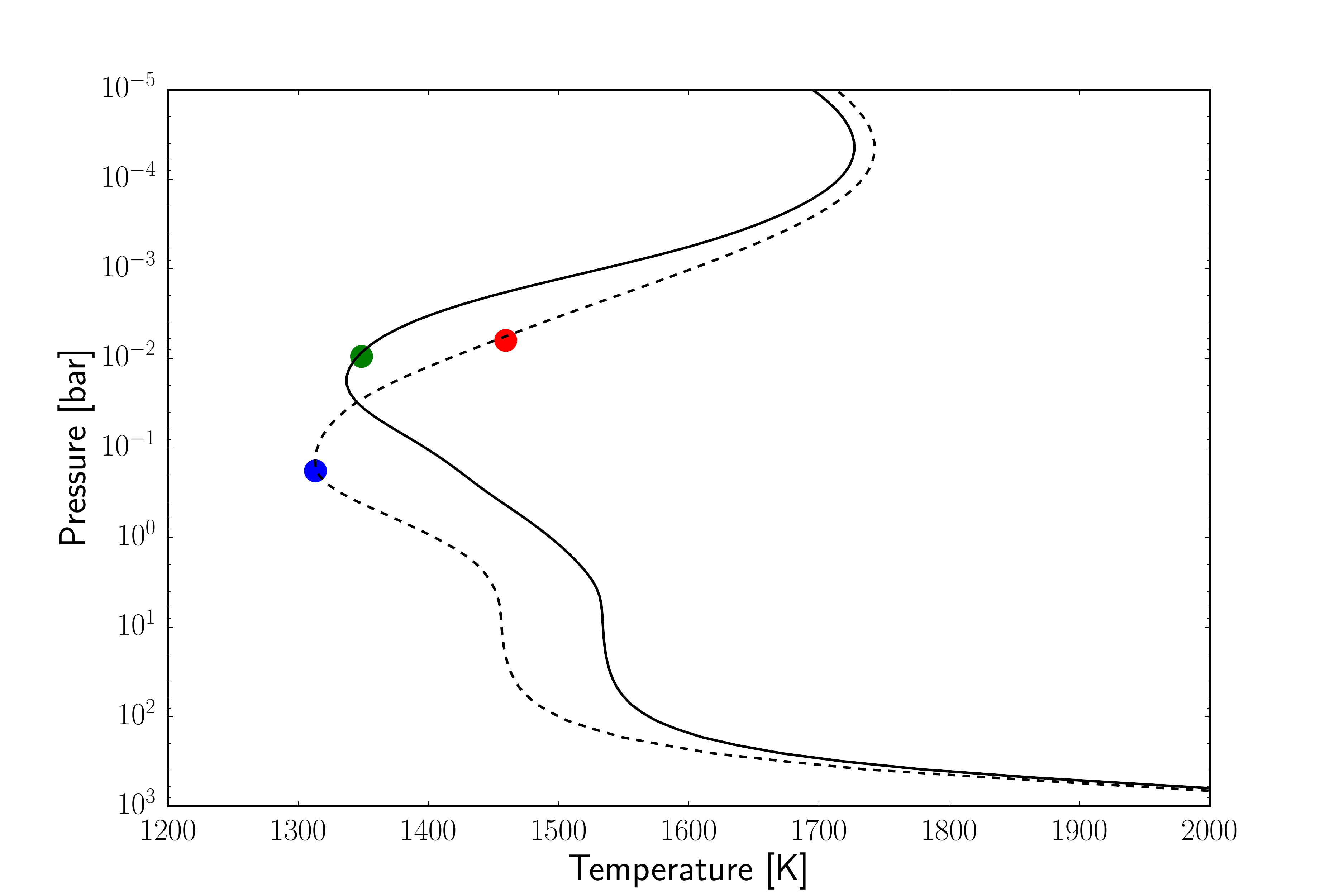}}
	\caption{The location of the peak emission in the 3.6 \textmu m Spitzer band for the HD~209458b model with a temperature inversion with the EQ calculation (blue), NEQ calculation (red) and CNEQ calculation (green); $K_{zz}$ = 10$^{11}$ cm$^2$s$^{-1}$.}
	\label{figure:3.6micron_hd209_tiovo} 
\end{figure}

\cref{figure:3.6micron_hd209_tiovo} indicates the pressure level of the photosphere in the 3.6 \textmu m Spitzer/IRAC band for the temperature inversion model. Similarly to the HD~189733b model, we see that transport-induced quenching pushes the photosphere to lower pressures by increasing the opacity. In this model however, where temperature is increasing with decreasing pressure, the photosphere moves to a higher temperature. This explains why we see an increased integrated flux in the NEQ model. However, for the CNEQ model, though the pressure level of the photosphere is still shifted to lower pressures the $P$--$T$ profile has adapted so that the temperature at this level is now cooler, conserving the integrated flux to that of the EQ model, shown in \cref{table:integrated_flux_hd209_tiovo}.

For the model without a temperature inversion, a similar process occurs. However, because temperature decreases with altitude in this model, the new photosphere is both lower in pressure \textit{and} lower in temperature resulting in a reduced emission flux. Therefore, to compensate, the $P$--$T$ structure increases in temperature at this pressure level to increase the emission flux once again, and conserve energy balance.

This understanding allows us to explain why the simulated spectra from our CNEQ models are similar to the EQ model spectra, reducing the impact of non-equilibrium chemistry. Our results show that it is not primarily the changing abundances due to non-equilibrium chemistry which effect the emission spectrum. It is instead the secondary effect of the non-equilibrium abundances shifting the location of the photosphere, by changing the opacity, which effects the calculated spectral flux, as the flux is now originating from a part of the atmosphere with a lower/higher temperature. This non-consistent method does not conserve energy balance in the atmosphere. 

On the other hand, in the self-consistent non-equilibrium chemistry models, though the non-equilibrium abundances do also change the pressure level of the photosphere, the $P$--$T$ profile adapts, by either heating up or cooling down, to maintain energy balance. Therefore, the resulting spectral flux for the CNEQ model shows a smaller discrepancy with the EQ model. Previous studies using non-consistent models may have overestimated the importance of non-equilibrium chemistry on the emission spectrum.

\section{Conclusions}
\label{section:conclusion}

We have presented results based on a fully-consistent chemical kinetics model applied to the atmospheres of HD~189733b and HD~209458b. 

Our simulations show that in cases of strong disequilibrium chemistry transport-induced quenching of absorbing species can induce changes in the $P$--$T$ profile of up to 100 K. These temperature shifts can, in turn, have impacts on both the chemical abundances themselves and on the corresponding emission spectra. The chemical abundances are affected via two related processes: firstly, temperature changes at high pressures, where the chemistry remains in chemical equilibrium, induce new chemical equilibrium abundances, and secondly, the change in temperature shifts the quenching point which alters the quenched abundances at lower pressures.

For instance, in our model of HD~189733b we would conclude that CH$_4$ is more abundant than CO for the $K_{zz}$ = 10$^{11}$ cm$^{2}$s$^{-1}$ case when not performing the calculation consistently. Instead, in the consistent approach, we find that CO is the most abundant carbon-bearing species. For the model of HD~209458b (with a temperature inversion) we find that the abundances of CH$_4$ and NH$_3$ are a factor of $\sim$5 and $\sim$3.5 lower in the consistent model compared with the non-consistent model, due to an increase in the temperature by more than 100 K in the deep atmosphere.

For both HD~189733b and HD~209458b models we find that consistent calculations of non-equilibrium chemistry reduce the overall impact of chemical disequilibrium on the emission spectrum. Our results show that in conventional chemical kinetics models, where the $P$--$T$ profile is held fixed, the dominant mechanism for non-equilibrium chemistry to affect the emission spectrum is by changing the pressure level, and temperature, of the photosphere. The strong dependence of the emission flux on temperature will result in a very different simulated emission spectrum.

However, in our consistent model the $P$--$T$ profile adapts to the new non-equilibrium chemical composition and to retain energy balance in the model atmosphere. The consequent temperature changes mitigate the effect of changing the location of the photosphere by either heating up or cooling down at the location of the new photosphere to preserve energy balance.

Based on these results, we urge caution when assessing the impact of non-equilibrium chemistry (transport-induced quenching and photochemistry) on the emission spectrum. Not including consistency between the chemical abundances and the temperature structure can lead to overestimates of the impact of non-equilibrium chemistry. 

This work has only consisdered 1D (vertical) effects of non-equilibrium chemistry. Horizontal advection is expected to be very important in the atmospheres of tidally-locked exoplanets which possess very strong zonal wind velocities \citep{Showman2009,Heng2011,RauscherMenou2012,Mayne2014}. In addition, these atmospheres can possess very large day-night temperature contrasts leading to large contrasts in horizontal chemical equilibrium abundances\citep{Burrows2010,Kataria2016}. Disequilibrium chemistry has already been suggested as a possible explanation to explain the discrepencies between the observed and model emission phase curves \citep{Zellem2014}. However, this work has shown that when performed consistently, transport-induced quenching has a smaller impact on the emission spectrum than previous studies suggest. It would therefore be very interesting and timely to study the process considered here including horizontal advection, by coupling a chemical kinetics scheme consistently to a 3D GCM.

\begin{acknowledgements}
This work is partly supported by the European
Research Council under the European Community’s Seventh Framework Programme
(FP7/2007-2013 Grant Agreement No. 247060-PEPS and grant No.
320478-TOFU). BD thanks the University of Exeter for support through a PhD studentship. DSA acknowledges support from the NASA Astrobiology Program through the Nexus for Exoplanet System Science. NJM and JG's contributions were in part funded by a Leverhulme Trust Research Project Grant, and in part by a University of Exeter College of Engineering, Mathematics and Physical Sciences studentship. This work used the DiRAC Complexity system, operated by the University of Leicester IT Services, which forms part of the STFC DiRAC HPC Facility. This equipment is funded by BIS National E-Infrastructure capital grant ST/K000373/1 and STFC DiRAC Operations grant ST/K0003259/1.
DiRAC is part of the National E-Infrastructure.
This work also used the University of Exeter Supercomputer, a DiRAC Facility jointly funded by STFC, the Large Facilities Capital Fund of BIS and the University
of Exeter.
\end{acknowledgements}

\bibliographystyle{aa} 

\begin{appendix}
\section{Model Testing}

In this appendix section we perform several comparisons with previously published works to benchmark both the radiative-convective and the chemistry modules of \texttt{ATMO}. 

\subsection{Radiative-Convective Scheme}
\label{section:radiative_convective_test}

\begin{figure}
	\centering
	\resizebox{\hsize}{!}{\includegraphics{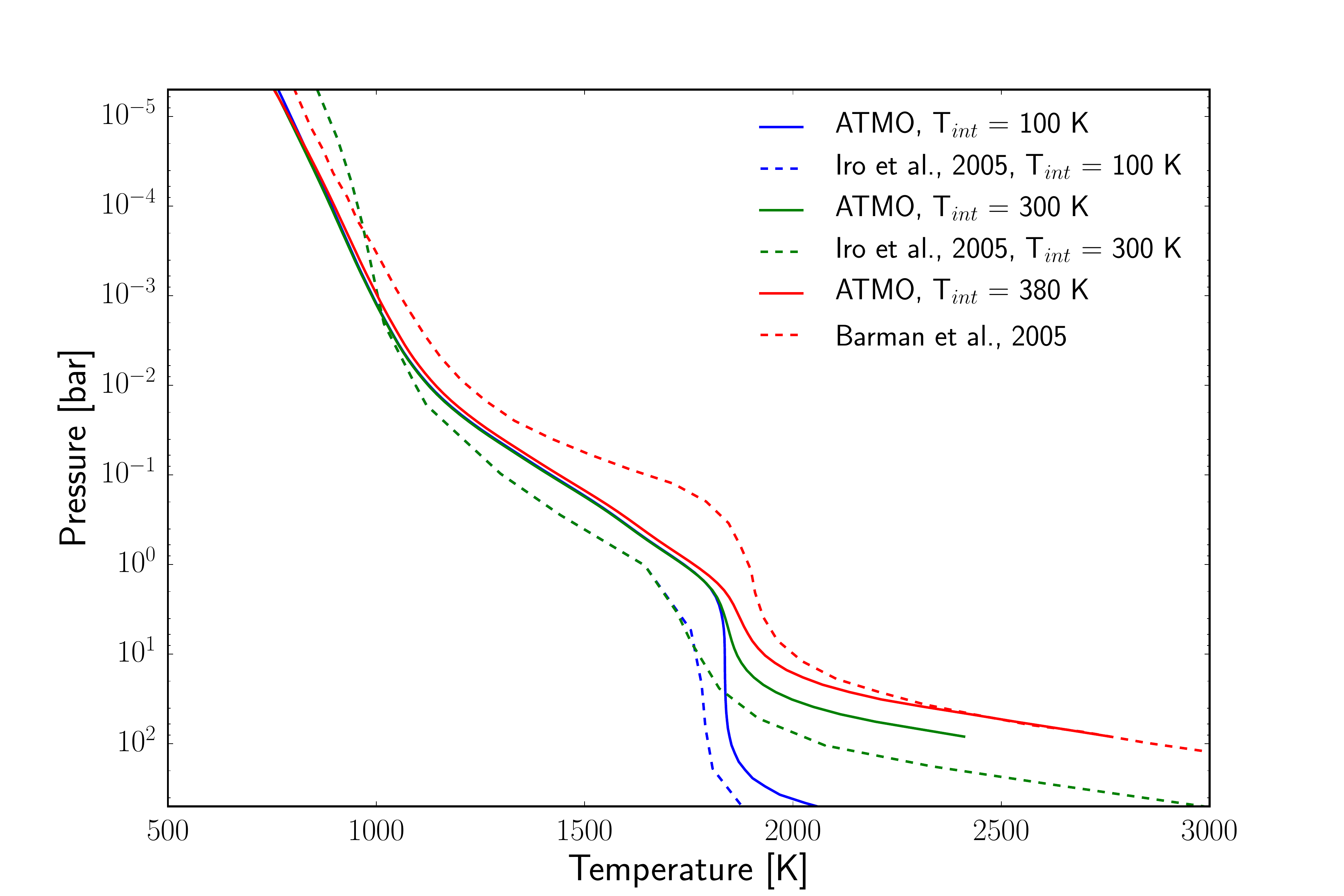}}
	\caption{A comparison of the $P$--$T$ profiles for HD~209458b with \texttt{ATMO} results in solid lines and the models of \citet{Iro2005} and \citet{Barman2005} in dashed lines.}
	\label{figure:iro_barman_comparison} 
\end{figure}

In this section we examine the radiative-convective equilibrium component of \texttt{ATMO} by reproducing the $P$--$T$ profiles of \citet{Iro2005} and \citet{Barman2005} (hereafter I05 and B05) for HD~209458b (\cref{figure:iro_barman_comparison}). In both cases we match the planetary parameters (i.e. internal temperature, gravity etc.) as closely as possible to those used in the original studies. Remaining differences between our model and those of I05 and B05 stem from different opacity sources, the stellar irradiation spectrum, and the radiative-transfer and chemistry schemes.

I05 models an atmoshere with an internal temperature, $T_{\rm int}$, of 100 K and 300 K and a planet-wide heat redistrubtion parameter; i.e. with an incident flux reduction factor of $\alpha$ = 0.25 which accounts for emission from both the dayside and nightside of the planet.

Generally our $P$--$T$ profile agrees very well with that of I05, though our model is $\sim$100K warmer throughout most of the modeled pressure range. This very likely stems from the different stellar irradiation used in the two models. I05 use a solar irradiation in their model, however we use the Kurucz simulated spectrum of HD~209458 which is several hundred Kelvin hotter than the Sun. Differences also likely stem from the treatment and source of opacities.

We also compare with the $\alpha$ = 0.25 model of B05. Note that we adjusted the internal temperature parameter until the profiles reached the same adiabat at depth since we do not know the exact value of $T_{\rm int}$ used in the B05 model. We found that $T_{\rm int}$=380 K matched the B05 profile well. We find that the \texttt{ATMO} profile is consistently lower in temperature with a maximum discrepancy at $\sim$0.1 bar. Aside from the differences in opacity sources (we use more up-to-date sources with increased infrared opacity)  between our model and the B05 model, the calculation of equilibrium chemistry is significantly different. B05 account for the 'rainout' process whereby condensation of species removes material from the gas-phase, depleting the elemental abundance of those species at lower pressures. We do not include this process in \texttt{ATMO}, potentially leading to differences in chemical composition and hence opacity.

The \texttt{ATMO} models of HD~209458b are bracketed by those of I05 and B05. There are significant differences in the opacity sources, irradiation, chemical and radiative schemes between all three models. Considering these significant differences, the three models generally agree on the qualitative temperature structure, with quantitative differences which should be expected due to the differences in model inputs and schemes.

\subsection{Chemistry Schemes}

\subsubsection{Gibbs Minimisation Scheme}
\label{section:chemical_gibbs_test}

\begin{figure}
	\centering
	\resizebox{\hsize}{!}{\includegraphics{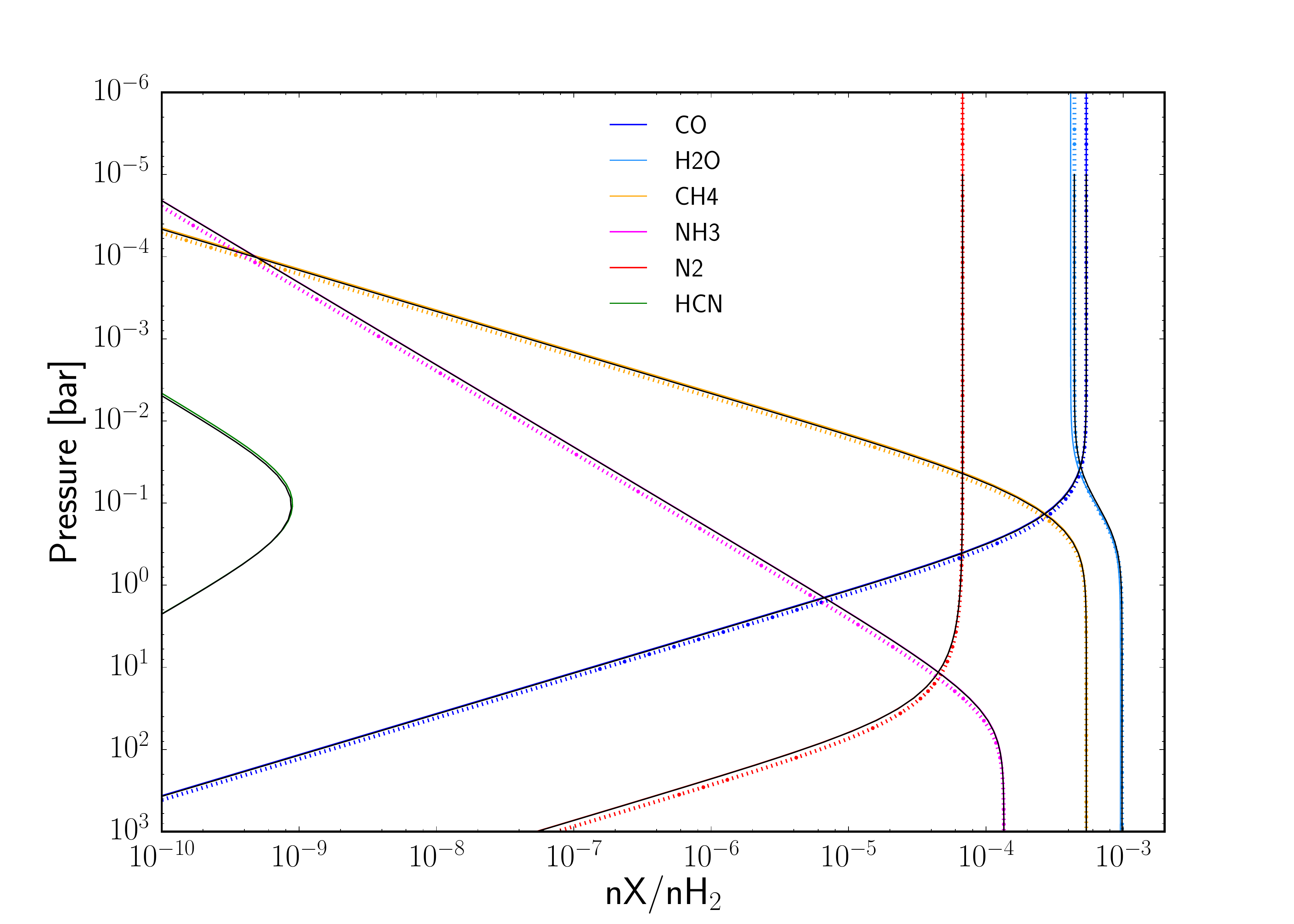}}
	\caption{The chemical equilibrium abundances derived using the Gibbs energy minimisation scheme (coloured solid lines), the \citet{Burrows1999} analytical formula (coloured dashed lines) and the \citet{Heng2016} analytical formula (black lines) for a $T$ = 1000 K isothermal profile. Note that HCN is not included in the \citet{Burrows1999} formula.}
	\label{figure:bs_heng_tcst1000} 
\end{figure}

\begin{figure}
	\centering
	\resizebox{\hsize}{!}{\includegraphics{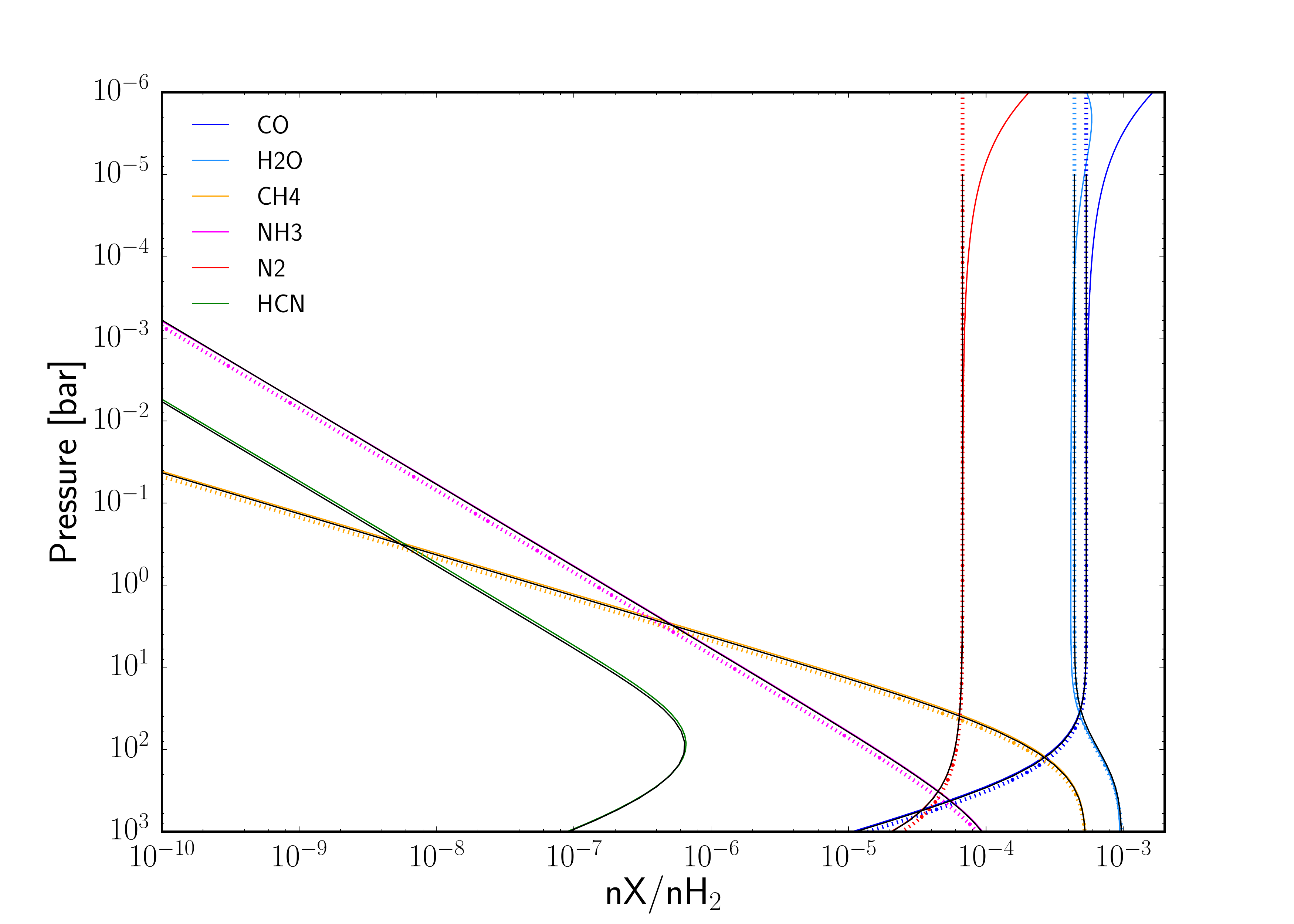}}
	\caption{As \cref{figure:bs_heng_tcst1000} for a $T$ = 2000 K isothermal profile.}
	\label{figure:bs_heng_tcst2000} 
\end{figure}

In this section we test the Gibbs minimisation scheme by comparing our chemical equilibrium abundances with those derived using the \citet{Burrows1999} and \citet{Heng2016} analytical formulae. \cref{figure:bs_heng_tcst1000,figure:bs_heng_tcst2000} show the abundances of CO, H$_2$O, CH$_4$, NH$_3$, N$_2$ and HCN derived using the Gibbs energy minimisation scheme and the two analytical methods for $T$ = 1000 K and 2000 K isothermal profiles, respectively; note that HCN is not included in the \citet{Burrows1999} scheme.

Our Gibbs minimisation scheme agrees very well with the analytical formula of \citet{Heng2016} for both temperature profiles. The agreement with the \citet{Burrows1999} formula is also very good. At low pressures in the $T$ = 2000 K model the Gibbs minimisation abundances deviate at low pressures due to molecular hydrogen dissociation which is not included in the analytical formula.

\subsubsection{Chemical Kinetics Scheme}

\label{section:chemical_kinetics_test}

In this section we test the chemical kinetics scheme implemented in our model.

\begin{figure*}
	\centering
	\resizebox{\hsize}{!}{\includegraphics{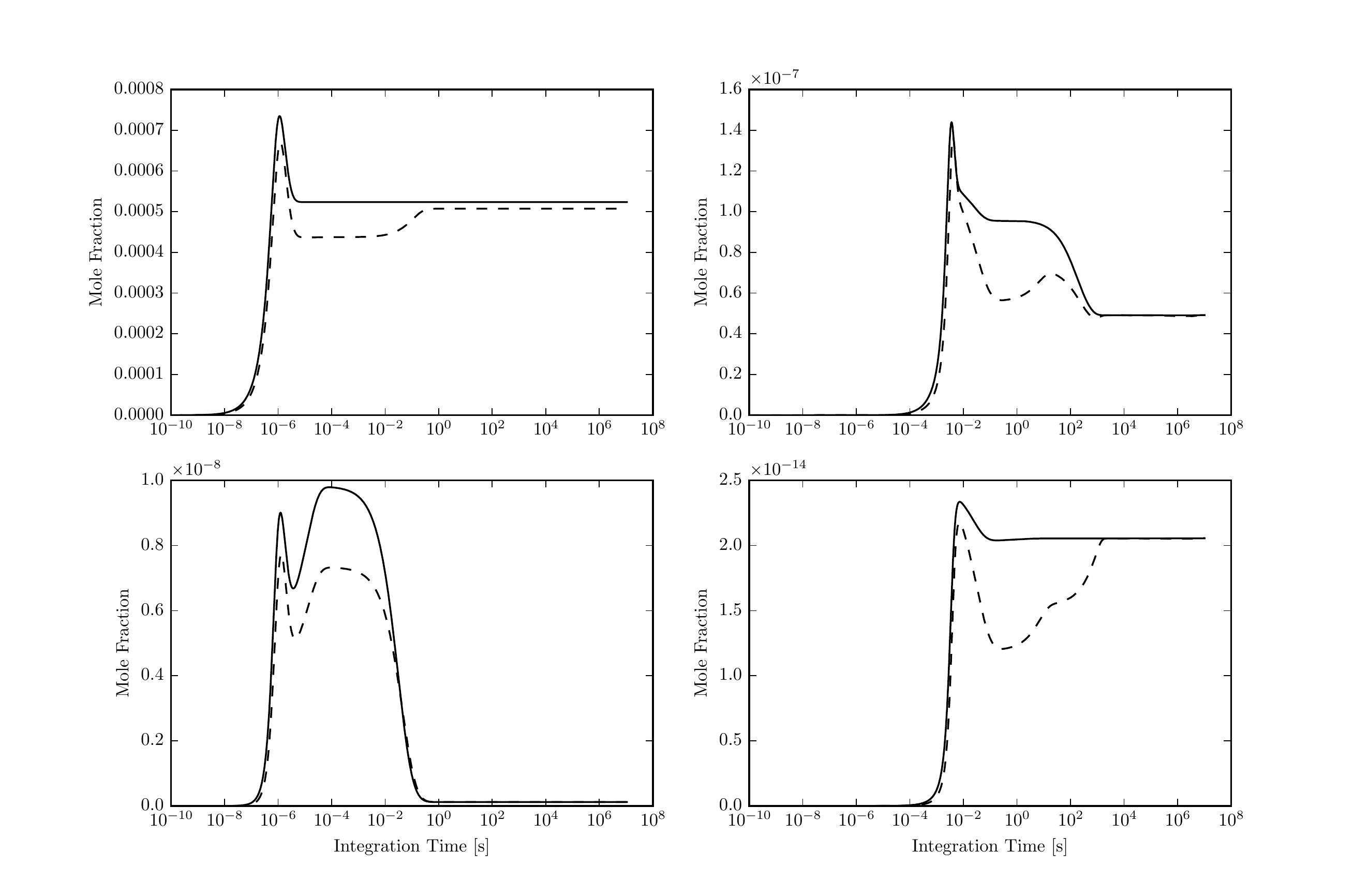}}
	\caption{Chemical evolution of (from top left, clockwise) H, OH, NNH and C$_2$H$_6$ with different approaches to the reversal of certain chemical reactions \citep[see][for details]{Venot2012} in the dashed and solid lines; to be compared with \citet[][Fig. 2]{Venot2012}.}
	\label{figure:chemical_evolution} 
\end{figure*}

Firstly we test our chemical kinetics scheme in isolation (i.e. only calculation of the chemical rates) using a 0D (box) model by reproducing the chemical evolution plots of \citet{Venot2012} (their figure 2) for H, OH, NNH and C$_2$H$_6$ (\cref{figure:chemical_evolution}). The evolution and final abundances of these species are, by eye, identical, validating the implementation of the \citet{Venot2012} chemical network in our model. 

Secondly, we reproduce the 1D chemical abundances profiles for HD~209458b included in \citet{Venot2012} using the same model inputs; pressure-temperature profile, chemical network, UV spectrum and eddy diffusion coefficient profile. \citet{Venot2012} used the same model inputs as \citet{Moses2011}, and compared the two different chemical networks. This allows us to compare with \citet{Venot2012}, whose chemical network we use here, testing only differences in the model implementation, and also with \citet{Moses2011}, who use both a different model and a different chemical network. The primary difference between the \texttt{ATMO} scheme and that of \citet{Venot2012} is the calculation of the UV flux. We calculate the UV flux as a function of pressure using the radiative transfer scheme as decribed in \cref{section:radiative_transfer}, whereas \citet{Venot2012} use the scheme of \citet{Isaksen1977}, which has a more simple treatment of scattering.

\begin{figure}
	\centering
	\resizebox{\hsize}{!}{\includegraphics{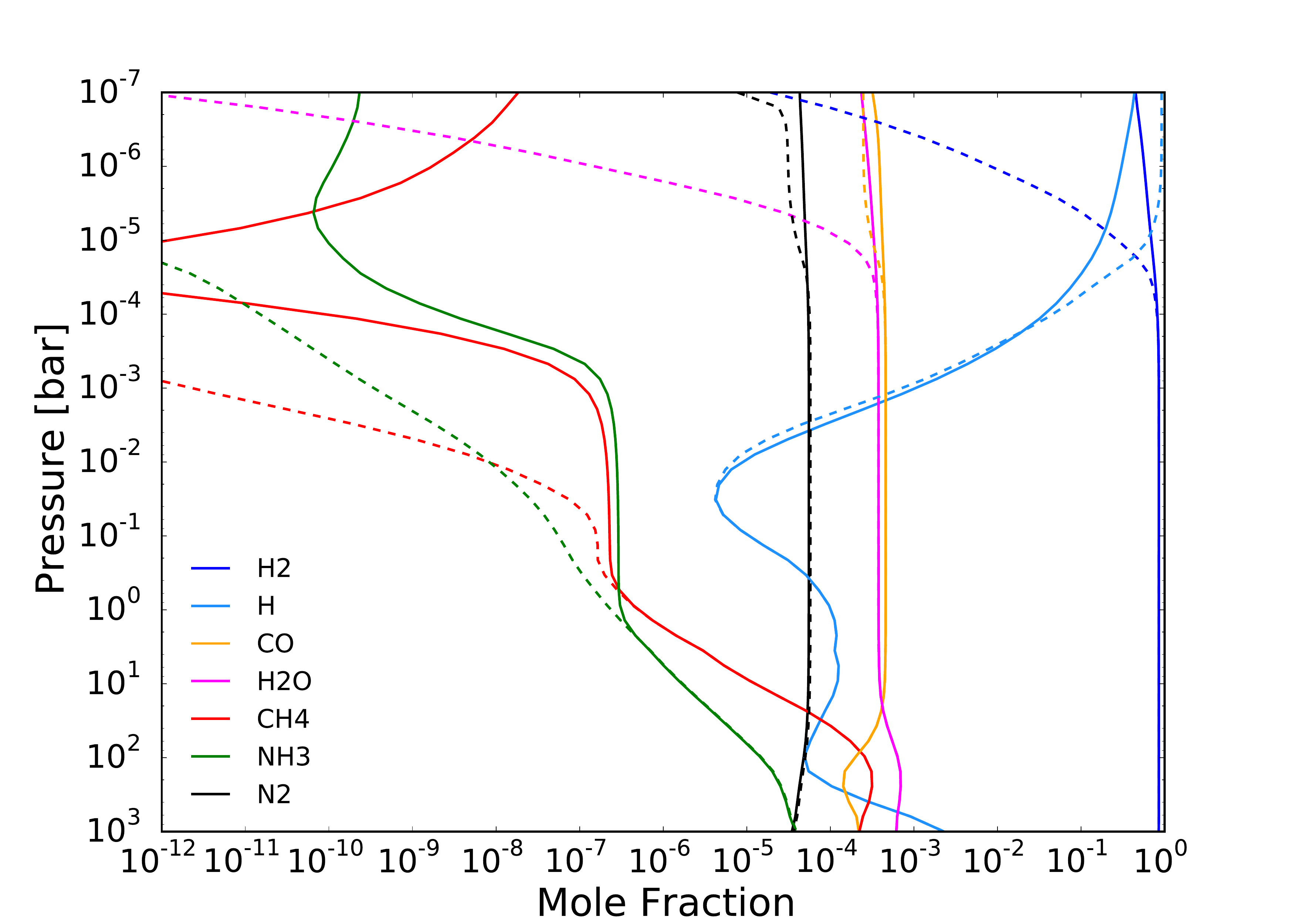}}
	\caption{The chemical abundances calculated for HD~209458b with equilibrium abundances (dashed) and non-equilibrium abundances (solid) using the \citet{Moses2011} pressure-temperature profile (their Fig. 2, dayside average profile); to be compared with \citet[][Fig. 5]{Venot2012} and \citet[][Fig. 5 and 6]{Moses2011}.}
	\label{figure:hd209_venot_comparison} 
\end{figure}

\cref{figure:hd209_venot_comparison} shows the chemical abundances calculated with \texttt{ATMO}, abundances calculated from the chemical kinetics scheme including vertical mixing and photochemistry. Overall, there is a very good agreement between our model and that of \citet{Venot2012}. In the deep atmosphere we retain chemical equilibrium, due to the very fast chemical timescale due to high pressures and temperatures. The abundances also agree very well in the middle regions of the atmosphere, with accurate reproductions of the methane and ammonia quenching points.

There are, however, some disagreements for the abundances at low pressures. As noted previously, one of the most important differences between our model and that of \citet{Venot2012} is the calculation of the UV flux. In particular, we find a production of methane at low pressures, when photochemistry is included, which is not seen in the \citet{Venot2012} model. Interestingly, this feature is present in the \citet{Moses2011} model (see their figure 5). The \citet{Moses2011} model uses Feautrier radiative-transfer method \citep{Michelangeli1992}, taking into account multiple-scattering by H$_2$ and He. \citet{Venot2012} note that abundances at low pressures are sensitive to the treatment of Rayleigh scattering.

In summary, our Gibbs minimisation scheme and chemical kinetics scheme, using the \citet{Venot2012} chemical network,  agree very well for chemical equilibrium. Our chemical equilibrium results agree well with \citet{Venot2012} and \citet{Moses2011}, and with the analytical formula of \citet{Burrows1999}. We also find a reasonably good agreement with \citet{Venot2012} when including vertical mixing and photochemistry, reproducing the quench points of the various species well. However, we find some discrepencies at low pressures where photochemistry is important, which we attribute to differences in the calculation of the UV flux.
 

\end{appendix}

\end{document}